\documentclass[preprint2]{emulateapj}

\newcommand{\chn}{{\it Chandra}}

\usepackage{graphicx}
\usepackage{longtable}
\usepackage{morefloats}

\slugcomment{version \today: fm}

\shorttitle{A \chn\ Snapshot Survey for 3C Radio Galaxies with 0.3$<$z$<$0.5}
\shortauthors{F. Massaro et al.  2013}

\begin{document}
\title{A \chn\ Snapshot Survey for 3C Radio Galaxies with redshifts between 0.3 and 0.5}

\author{
F. Massaro\altaffilmark{1}, 
D. E. Harris\altaffilmark{2},
G. R. Tremblay\altaffilmark{3},
E. Liuzzo\altaffilmark{4},
A. Bonafede\altaffilmark{5},
A. Paggi\altaffilmark{2}.
}

\altaffiltext{1}{SLAC National Laboratory and Kavli Institute for Particle Astrophysics and Cosmology, 2575 Sand Hill Road, Menlo Park, CA 94025, USA.}
\altaffiltext{2}{Smithsonian Astrophysical Observatory, 60 Garden Street, Cambridge, MA 02138, USA.}
\altaffiltext{3}{European Southern Observatory, Karl-Schwarzschild-Str. 2, 85748 Garching bei Muenchen, Germany.}
\altaffiltext{4}{Istituto di Radioastronomia, INAF, via Gobetti 101, 40129, Bologna, Italy.}
\altaffiltext{5}{Hamburger Sternwarte, Universit$\ddot{a}$t Hamburg, Gojenbergsweg 112, 21029 Hamburg, Germany.}

\begin{abstract} 
This paper contains an analysis of short \chn\ observations of 19 3C
sources with redshifts between 0.3 and 0.5 not previously observed in
the X-rays.  This sample is part of a project to obtain \chn\ data for
all of the extragalactic sources in the 3C catalogue.  Nuclear X-ray
intensities as well as any X-ray emission associated with radio jet
knots, hotspots or lobes have been measured in 3 energy bands: soft,
medium and hard.  Standard X-ray spectral analysis for the 4 brightest
nuclei has been also performed.  X-ray emission was detected for all
the nuclei of the radio sources in the current sample with the
exception of 3C\,435A.  There is one compact steep spectrum (CSS)
source while all the others are FR\,II radio galaxies. X-ray emission
from two galaxy clusters (3C\,19 and 3C\,320); from 6 hotspots in 4
radio galaxies (3C\,16, 3C\,19, 3C\,268.2, 3C\,313); and extended
X-ray emission on kpc scales in 3C\,187 and 3C\,313, has been
detected.
\end{abstract}


\section{Introduction}
\label{sec:intro}
The revised version of the Third Cambridge (3C) catalog presented in
Spinard et al. (1985) lists 298 extragalactic radio sources \citep[see
 also][for additional information]{edge59,mackay71}.  In recent years, the
large majority of these sources has been observed with several
photometric and spectroscopic snapshot surveys approaching the
statistical completeness of the 3C (radio flux limited) sample.  
The Hubble Space Telescope (HST) already
completed an optical snapshot survey of the 3C sources at redshift
lower than 0.3 \citep[e.g.,][and reference
 therein]{chiaberge00,tremblay09} while ground based spectroscopic
observations have been carried out with the Galileo Telescope
\citep{buttiglione09}.  In addition, radio images with arcsecond
resolution for the majority of the 3C sources are available
from the NRAO VLA Archive Survey (NVAS)\footnote{http://www.aoc.nrao.edu/$\sim$vlbacald/}
and in the archives of the Very Large Array (VLA) and MERLIN observatories.  
Recently, a piecemeal approach of obtaining \chn\ observations for all the 3C
extragalactic sources below redshift 0.3 was completed \citep[][hereinafter Paper1 and Paper2, respectively]{massaro10,massaro12}. 
\chn\ is the only X-ray facility with angular resolution comparable to that at optical and radio frequencies.

In this paper we provide source parameters for 19 previously
unobserved (by \chn) 3C sources with redshifts between 0.3 and 0.5.
{ With these data, the Chandra coverage of 3CR sources with z$<$0.5 is
complete except for the following:  one quasar, 3 unclassified sources,
and 1 FRII radio galaxy which had no redshift listed in Spinrad et al. (1985).  We
plan to include all of these in our next Chandra Snapshot proposal.}
Our eventual goal is to have observations of all 3C extragalactic
sources in the \chn\ archives.
As has been amply demonstrated by the observations of sub samples (Cycle 9 and
Cycle 12) even with relatively short exposures, almost all
nuclei and occasionally X-ray emission associated with jet knots,
hotspots, lobes, and hot thermal gas (i.e. ICM of groups or clusters
of galaxies) have been clearly detected.

{ The whole 3CR sample \citep{spinrad85} contains 158 FR II radio galaxies,
39 FR I radio galaxies, 57 quasars, 2 Seyfert galaxies, 2 BL Lacs, 20
unidentified, and 20 unclassified sources.  Of these, 74 are still
unobserved by Chandra and to the best of our knowledge, 26 have no
redshift (see Appendix~\ref{sec:appendixA} for specifics).}

In Paper I, data reduction
and analysis procedures, developed for this project, were presented together with the
results for the first 30 radio sources.
In Paper II, a similar investigation of
an additional 27 sources, to complete the sample to $z\leq$0.3 has been carried out.  

The paper is organized as follows: a brief description of the
observations and data reduction procedures is given in
\S~\ref{sec:obs} while source details are given in
\S~\ref{sec:results}.  Then, \S~\ref{sec:summary} is devoted to our
summary and conclusions.

For numerical results, cgs units are used unless stated otherwise
and a flat cosmology was assumed with $H_0=72$ km s$^{-1}$ Mpc$^{-1}$,
$\Omega_{M}=0.27$ and $\Omega_{\Lambda}=0.73$ \citep{dunkley09}.
Spectral indices, $\alpha$, are defined by flux density,
S$_{\nu}\propto\nu^{-\alpha}$.

\section{Observations, Data Reduction, and Basic Parameters}
\label{sec:obs}
Data reduction and data analysis procedures have been extensively described 
in Paper I and II, thus only the basic details are reported here.

The 3C source sample observed during Cycle 13 is listed in
Table~\ref{tab:main} together with their salient parameters.  Each
source was observed for a nominal 12 ksec, and the actual livetimes
are given in Table~\ref{tab:fluxes}, together with the nuclear X-ray
fluxes.  It is worth noting that in Cycle 13 the \chn\ exposure time
per single source has been increased by 50\% with respect to the
previous 3C snapshot survey carried out in Cycles 9 and 12 observed
with 8 ksec nominal exposure.  The ACIS-S back illuminated chip in
VERY FAINT mode with standard frame times (3.2s) has been used.  All
the observations had 4 chips turned on: I2, I3, S2, and S3.  The data
reduction has been performed following the standard reduction
procedure described in the \chn\ Interactive Analysis of Observations
(CIAO)
threads\footnote{http://cxc.harvard.edu/ciao/guides/index.html}, using
CIAO v4.4 and the \chn\ Calibration Database (CALDB) version 4.5.3.

Level 2 event files were generated using the $acis\_process\_events$
task and events were filtered for grades 0,2,3,4,6. 
Lightcurves were also extracted for every dataset to verify the absence of high background
intervals.  Astrometric registration was achieved by changing the
appropriate keywords in the fits header so as to align the nuclear
X-ray position with that of the radio (see Section 2 in Paper II).

\begin{table*} 
\caption{Source List of the \chn\ Cycle 13 Snapshot Survey of  3C Radio Sources with 0.3$<$z$<$0.5}
\label{tab:main}
\begin{tabular}{lllllrcccrcl}
\hline
3C     & Class\tablenotemark{(a)}  & R.A. (J2000)\tablenotemark{(b)}   & Dec. (J2000)\tablenotemark{(b)}  & z\tablenotemark{(c)}       & D$_L$
          & Scale       & N$_{H,Gal}$\tablenotemark{(d)}    & m$_v$\tablenotemark{(e)} & S$_{178}$\tablenotemark{(f)}  & \chn\    & Obs. Date \\
          &                     & hh mm ss      & dd mm ss         &        &  (Mpc)   & (kpc/\arcsec)      & (cm$^{-2}$)&     & (Jy)                 & Obs ID & yyyy-mm-dd \\ 
\hline 
\noalign{\smallskip}
~~16    &  FR II - HEG & 00 37 45.40 & +13 20 09.2 & 0.405  & 2161.8 & 5.309       & 4.58e20 & 21.0 & 11.2 & 13879  & 2012-10-25 \\ 
~~19    &  FR II - LEG & 00 40 54.99 & +33 10 07.1 & 0.482  & 2662.9 & 5.878       & 6.18e20 & 20.0 & 12.1 & 13880  & 2011-10-29 \\ 
~~42    &  FR II - HEG & 01 28 30.12 & +29 03 00.8 & 0.395  & 2098.5 & 5.228       & 6.42e20 & 20.0 & 12.0 & 13872  & 2012-02-26 \\
~~46    &  FR II - HEG & 01 35 28.48 & +37 54 05.2 & 0.437 & 2369.1 & 5.560       & 5.64e20 & 19.5 & 10.2 & 13881  & 2012-09-28 \\ 
~~67    &  CSS - BLO   & 02 24 12.29 & +27 50 11.5 & 0.310 & 1579.7 & 4.462       & 7.47e20 & 18.0 & 10.0 & 13873  & 2011-11-27 \\ 
103     &  FR II - ?   & 04 08 03.10 & +43 00 32.7 & 0.330   & 1697.9 & 4.654       & 3.18e21 & 19.0 & 26.6 & 13874  & 2012-11-10 \\ %
187     &  FR II - ?   & 07 45 04.46 & +02 00 08.7 & 0.350   & 1819.2 & 4.839       & 6.00e20 & 19.5 &  8.1 & 13875  & 2012-01-04 \\ 
244.1   &  FR II - HEG & 10 33 33.98 & +58 14 35.7 & 0.428  & 2308.9 & 5.489       & 6.67e19 & 19.0 & 20.3 & 13882  & 2013-01-17 \\ 
268.2   &  FR II - ?   & 12 00 58.61 & +31 33 19.9  & 0.362  & 1892.8 & 4.947       & 1.64e20 & 19.0 &  9.7 & 13876  & 2012-07-07 \\ 
274.1   &  FR II - ?   & 12 35 26.66 & +21 20 34.8 & 0.422  & 2270.3 & 5.443       & 2.35e20 & 20.0 & 16.5 & 13883  & 2011-11-06 \\
275     &  FR II - LEG & 12 42 19.89 & -04 46 20.1 & 0.480   & 2649.6 & 5.864       & 1.86e20 & 21.0 & 14.5 & 13884  & 2012-07-12 \\ 
306.1   &  FR II - HEG & 14 55 01.40 & -04 20 59.8 & 0.441  & 2393.1 & 5.587       & 6.11e20 & 19.0 & 13.5 & 13885  & 2012-09-06 \\ 
313     &  FR II - HEG & 15 11 00.03 & +07 51 50.1 & 0.461  & 2523.9 & 5.732       & 2.40e20 & 21.0 & 20.6 & 13886  & 2012-05-07 \\
320     &  FR II - ?   & 15 31 25.37 & +35 33 40.0 & 0.342  & 1770.5 & 4.766       & 1.69e20 & 18.0 &  9.1 & 13877  & 2011-12-02 \\ 
327.1   &  ? - HEG     & 16 04 45.38 & +01 17 50.3 & 0.462  & 2530.4 & 5.740       & 6.40e20 & 20.5 & 23.6 & 13887  & 2012-05-08 \\
341     &  FR II - HEG & 16 28 03.98 & +27 41 39.3 & 0.448  & 2438.7 & 5.639       & 3.51e20 & 19.0 & 10.8 & 13888  & 2011-11-14 \\
411     &  FR II - HEG & 20 22 08.44 & +10 01 11.3 & 0.467  & 2563.4 & 5.775       & 1.06e21 & 19.7 & 16.5 & 13889  & 2012-08-08 \\
434     &  FR II - ?   & 21 23 16.24 & +15 48 05.8 & 0.322  & 1649.9 & 4.577       & 6.57e20 & 20.8 &  4.8 & 13878  & 2012-08-15 \\ 
435B     &  FR II - ?  & 21 29 06.10 & +07 32 54.8 & 0.865  & 5460.0 & 7.610       & 4.44e20 & 19.4 & --- & 13890  & 2012-08-14 \\
\noalign{\smallskip}
\hline
\end{tabular}\\
(a) The `class' column contains both a radio descriptor (Fanaroff-Riley class I or II), Compact Steep Spectrum (CSS) and 
the optical spectroscopic designation, LEG, ``Low Excitation Galaxy'', HEG, ``High Excitation Galaxy'', and BLO, ``Broad Line Object''.
The symbol ``?" indicates those radio and optical classifications that are uncertain or not reported in the literature.\\
(b) The celestial positions listed are those of the radio nuclei which we used to register the X-ray images except for\\
the 4 sources lacking an obvious radio nucleus.  For these four (3C\,16, 3C\,19, 3C\,268.2, and 3C\,275) we use Spinrad's\\
position \citep{spinrad85}.  For 3C\,275 the listed position falls a few arcsec E of the radio and X-ray emission.\\
(c) Redshift measurements are taken form Spinrad et al. (1985) or from NASA/IPAC Extragalactic Database (NED) for 3C\,435B.\\
(d) Galactic Neutral hydrogen column densities N$_{H,Gal}$ are taken form Kalberla et al. (2005).\\
(e) $m_v$ is the visual magnitude (Spinrad et al. 1985) or from NASA/IPAC Extragalactic Database (NED) for 3C\,435B.\\
(f) S$_{178}$ is the flux density at 178 MHz, taken from Spinrad et al. (1985).\\
\end{table*}

\subsection{Fluxmaps}
\label{sec:fluxmaps}
Three different fluxmaps were created in the energy ranges: 0.5 -- 1 keV
(soft), 1 -- 2 keV (medium), 2 -- 7 keV (hard), by
filtering the event file with the appropriate energy range and
dividing the data with monochromatic exposure maps (with nominal
energies of soft=0.8keV, medium=1.4keV, and hard=4keV).  

To recover the angular resolution of the Chandra mirrors,
the undersampling imposed by the ACIS pixel size was avoided by regridding to
obtain pixel sizes of 0.123\arcsec\ or smaller. 
For sources of large angular extent 1/2 or no regridding was used.

To obtain maps with brightness units of
ergs~cm$^{-2}$~s$^{-1}$~pixel$^{-1}$, each event was multiplied by the
nominal energy of its respective band { (see Section 2.1 of Paper II for additional details)}.

To measure observed fluxes for any feature, an appropriate region
(usually circular) was chosen.  For the nuclei, regions of
2\arcsec\ radius have been used while for other features the size is
given in Table~\ref{tab:jets}.  For each feature, background circular
regions, with the same size described above, were chosen so as to
avoid contaminating X-ray emission (and also radio emission) and to
sample both sides of jet features or two areas close to hotspots.
X-ray fluxes measured for the nuclei are given in
Table~\ref{tab:fluxes} while those of the hotspots and knots in
Table~\ref{tab:jets}.

\begin{table*}
\caption{Nuclear X-ray Fluxes}
\label{tab:fluxes}
\begin{tabular}{llrrrrrrrrrr}
\hline
3C     & LivTim\tablenotemark{(a)}&  Net\tablenotemark{(b)}& Ext. Ratio\tablenotemark{(c)}& f(soft)          & f(medium)& f(hard)    & f(total)        & HR & N$_H\left(z\right)\tablenotemark{(d)}$ & L$_X$ \\ 
          & (ksec)                                 & (cnts)                             &                                                   & 0.5-1~keV & 1-2~keV   & 2-7~keV & 0.5-7~keV &                                      & (10$^{22}$cm$^{-2}$)  & (10$^{42}$erg~s$^{-1}$) \\
\hline 
\noalign{\smallskip}
~~16     & 11.92   &  2.8(1.7)   &  0.22(0.15)   & 0.34(0.34)  &  ---    & 3.3(2.3)        & 3.6(2.3)        &  ---        & ---       &    2.0(1.3)       \\    
~~19     & 11.89   &   35(6)     &  0.16(0.03)   & ---        & ---        & 24.1(5.6)  & 24.1(5.6)  &  $<$1.00    & $<$36     &   20.6(4.7)  \\
~~42     & 11.92   &   39(6)     &  0.78(0.17)   & ---        & 0.56(0.56) & 46.6(7.8)  & 47.2(7.8)  &  0.98(0.23) & 4.7 - 25  &   24.8(4.1)  \\
~~46     & 11.92   &   50(7)     &  0.81(0.15)   & 0.79(0.46) & 0.88(0.51) & 63.5(10.0) & 65.2(10.0) &  0.97(0.22) & 2.8 - 36  &   43.8(6.7)  \\
~~67     & 11.89   &  567(24)    &  0.96(0.06)   & 30.3(3.2)  & 74.7(4.8)  & 257(17)    & 361(18)    &  0.55(0.06) & 1.9 - 5.0 &  108(5)      \\
103      & 11.92   &  241(15)    &  0.92(0.08)   & 0.36(0.36) &  6.3(1.5)  & 320(22)    & 327(22)    &  0.96(0.09) & 6.1 - 25  &  113(8)      \\
187      & 11.90   &    9(3)     &  0.15(0.05)   & 0.93(0.53) & ---        & 3.1(2.8)   & 4.0(2.9)   &  $<$1.00    & $<$25     &    1.6(1.2)  \\
244.1    & 11.92   &   44(7)     & 0.60(0.12)    & 3.9(1.2)   & 1.2(0.7)   & 39.5(7.5)  & 44.6(7.6)  &  0.94(0.25) & 4.3 - 36  & 28.4(4.8)    \\
268.2    & 11.58   &   12(3)     &  0.41(0.14)   & 0.55(0.39) & 1.3(0.7)   & 6.7(3.4)   & 8.6(3.5)   &  0.67(0.51) & $<$25     &    3.7(1.5)  \\
274.1    & 11.90   &   22(4)     &  0.49(0.12)   & ---        & 0.52(0.37) & 27.3(6.5)  & 27.8(6.5)  &  0.96(0.33) & 3.5 - 36  &   17.2(4.0)  \\
275      & 12.41   &   45(6)     &  0.80(0.16)   & 0.88(0.51) & 4.0(1.1)   & 30.1(5.8)  & 35.0(5.9)  &  0.77(0.22) & 3.0 - 29.4&   29.4(5.0)  \\
306.1    & 11.92   &   24(5)     &  0.63(0.16)   & 0.55(0.39) & ---        & 27.9(6.3)  & 28.5(6.3)  &  $<$1.00    & $<$36     &   19.5(4.3)  \\
313      & 11.92   &   24(5)     &  0.57(0.15)   & ---        & ---        & 33.7(6.7)  & 33.7(6.7)  &  $<$1.00    & $<$37     &   25.7(5.3)  \\
320      & 11.90   &   68(8)     &  0.13(0.02)   & 6.1(1.7)   & 5.3(1.7)   & 14.9(5.3)  & 26.3(5.8)  &  0.48(0.30) & $<$7.3    &    9.9(2.2)  \\
327.1\tablenotemark{(e)}    & 11.11   & 1235(35)    &  0.90(0.04)   & 96(10)     & 204(20)    & 935(94)    & 1236(97)   &  ---        & ---       &  946(74)     \\   
341      & 11.64   &   18(4)     &  0.64(0.19)   & 0.59(0.59) & 1.46(0.73) & 13.9(4.2)  & 16.0(4.3)  &  0.81(0.36) & 2.0 - 36  &   11.3(3.1)  \\
411\tablenotemark{(e)} & 11.91 & 1716(41) & 0.93(0.03) & 116(11) & 279(31) & 1627(155)  & 2023(158)  &  ---        & ---       &  1590(120)   \\
434      & 11.92   &   44(7)     &  0.83(0.17)   & 4.2(1.2)   & 6.1(1.4)   & 15.2(4.4)  & 25.5(4.8)  &  0.43(0.24) & $<$5.7    &  8.3(1.6)    \\
435B     & 11.92   &  293(17)    &  0.90(0.07)   & 5.6(1.4)   & 38.0(3.5)  & 179(14)    & 223(15)    &  0.65(0.08) & 3.2 - 8.4 &  795(43)     \\
\noalign{\smallskip}
\hline
\end{tabular}\\
\tablecomments{Fluxes are given in units of 10$^{-15}$erg~cm$^{-2}$s$^{-1}$.
Values in parentheses are 1$\sigma$ uncertainties.}

\tablenotemark{a}{LivTim is the live time} 

\tablenotemark{b}{Net is the net counts within a circle of radius=2\arcsec.}

\tablenotemark{c}{Ext. Ratio (``Extent Ratio'') is the ratio of the
net counts in the r\,=\,2\arcsec\ circle to the net counts in the
r\,=\,10\arcsec\ circle.  Values significantly less than 0.9
indicate the presence of extended emission around the nuclear
component { (see Section~\ref{sec:general})}.}  

\tablenotemark{d}{As per the discussion in the text, the
value of N$_H\left(z\right)$, required to produce the observed $HR$ values, was computed.  
The uncertainty given here is indicative only: it is the range
of N$_H\left(z\right)$ covered by the uncertainty in the $HR$ and allowing
$\alpha_X$ to range from 0.5 to 1.5.  Obviously there may be some
sources with intrinsic spectral indices outside of this range.}

\tablenotemark{e}{For 3C\,327.1 and 3C\,411, X-ray fluxes and luminosities
have been computed via spectral fitting including the jdpileup model (see \S~\ref{sec:spectra} for more details),
thus the N$_H\left(z\right)$ values are reported in Table~\ref{tab:spec}.
The `light bucket' estimates of the total flux (\S~2.2) are in reasonable agreement with those in the table: 1050 (3C327.1) and 2120 (3C411).}

\end{table*}

\begin{table*} 
\caption{Radio components with X-ray Detections}
\label{tab:jets}
\begin{center}
\begin{tabular}{rrrrrrrrrr}
\hline
3C                     &Component\tablenotemark{a}& Radius\tablenotemark{b}&counts\,(bkg)\tablenotemark{c}& Detection &   
f$_{0.5-1~keV}$&f$_{1-2~keV}$                        &f$_{2-7~keV}$                &f$_{0.5-7~keV}$                       &L$_X$\\ 
                          &                                               & (arcsec)                         &                                                 & Significance\tablenotemark{d} &
(cgs)                  &(cgs)                                       &(cgs)                               &(cgs)                                         &10$^{42}$erg~s$^{-1}$\\
\hline 
\noalign{\smallskip}
 ~~16    & $l$ - n\,11.1  & 3.0 &  5(2) & 2.1$\sigma$     &  ---  &  0.44(0.44)  &  1.37(1.37)  &  1.81(1.44)  &  1.01(0.80) \\
         & $l$ - n\,5.2 & 2.5 &  3(1) & 2.0$\sigma$	 &  0.36(0.36)  &  0.22(0.22)  &  ---  &  0.58(0.42)  &  0.32(0.24) \\
         & $l$ - s\,6.0 & 3.0 &  2(1) & 1.4$\sigma$	 &  0.37(0.37)  &  0.12(0.27)  &  ---  &  0.50(0.46)  &  0.28(0.26) \\
         & $h$ - s\,22.8 & 3.0 &  4(1) & 2.7$\sigma$	 &  0.40(0.40)  &  0.49(0.35)  &  ---  &  0.89(0.53)  &  0.50(0.30) \\
 ~~19    & $h$ - n\,3.0  & 0.8 &  5(1) & 3.2$\sigma$ 	 &  0.39(0.39)  &  0.73(0.53)  &  0.93(0.93)  &  2.05(1.14)  &  1.74(0.96) \\
         & $h$ - s\,3.5  & 0.8 &  6(2) & 2.5$\sigma$	 &  0.88(0.51)  &  0.78(0.67)  &  1.74(1.23)  &  3.40(1.49)  &  2.88(1.26) \\
  187    & $l$ - n\,37.0 & 12.0& 46(11)& $>$7.0$\sigma$	 &  4.02(1.30)  &  5.25(1.43)  &  2.37(2.37)  & 11.64(3.06)  &  4.61(1.21) \\
         & $l$ - n\,75.0 & 12.0& 30(5) & $>$7.0$\sigma$	 &  1.27(0.69)  &  1.16(0.63)  &  7.98(3.62)  & 10.41(3.74)  &  4.12(1.48) \\
         & $l$ - s\,27.0 & 12.0& 46(13)& $>$7.0$\sigma$	 &  2.86(1.31)  &  5.41(1.42)  &  7.19(4.64)  & 15.46(5.03)  &  6.12(1.99) \\
         & $l$ - s\,60.0 & 15.0& 56(15)& $>$7.0$\sigma$	 &  1.57(0.70)  &  2.51(1.06)  &  2.49(2.49)  &  6.57(2.80)  &  2.60(1.11) \\
  268.2  & $h$ - s\,46.0 & 2.0 &  2(1) & 1.4$\sigma$	 &  0.38(0.38)  &  0.14(0.14)  &  ---  &  0.52(0.40)  &  0.22(0.17) \\
  313    & $h$ - e\,73.0 & 1.5 &  4(1) & 2.7$\sigma$	 &  ---  &  0.53(0.37)  &  2.26(1.59)  &  2.79(1.63)  &  2.13(1.24) \\
         & $h$ - w\,55.0 & 1.5 &  5(1) & 3.2$\sigma$	 &  0.76(0.54)  &  0.22(0.22)  &  1.67(1.67)  &  2.64(1.77)  &  2.02(1.35) \\
  327.1  & $k$ - s\,3.3  &2.7x1&  9(3) & 3.0$\sigma$	 &  0.01(0.01)  &  2.17(0.90)  &  ---  &  2.18(0.90)  &  1.67(0.69) \\
         & $k$ - s\,5.0  & 0.8 &  3(1) & 2.0$\sigma$	 &  ---  &  0.69(0.49)  &  ---  &  0.69(0.49)  &  0.53(0.38) \\
  341    & $k$ - w\,8.0  & 1.5 &  3(1) & 2.0$\sigma$	 &  0.78(0.55)  &  ---  &  0.98(0.98)  &  1.76(1.12)  &  1.25(0.80) \\
\noalign{\smallskip}
\hline
\end{tabular}\\
\end{center}
Fluxes are given in units of 10$^{-15}$erg~cm$^{-2}$s$^{-1}$.\\
(a) The component designation is comprised 
of a letter indicating the classification (i.e., knot $k$, hotspot $h$, lobe $l$), 
a cardinal direction (as viewed from the nucleus) plus the distance from the nucleus in arcseconds.\\
(b) The radius column gives the size of the aperture used for photometry.\\
(c) The counts column gives the total counts in the photometric circle together 
with those measured in the background regions of the same area, in parentheses; both for the 0.5 to 7 keV band.\\
{ (d) The confidence level of each detection evaluated adopting a Poisson distribution.}\\
\end{table*}

\subsection{Fluxes for the two nuclei affected by pileup}
Two of the detected nuclei had count rates exceeding the adopted
pileup threshold of 0.2 counts per frame { (see Section 2.3 in Paper II)}.  These are 3C\,327.1 (0.45
counts/frame) and 3C\,411 (0.75 counts/frame).  Since pileup serves to move
events to higher energies, it was not possible to use fluxmaps to measure fluxes in
the 3 bands.  Instead, the jdpileup spectral fitting was adopted
(\S~2.3) to determine the appropriate fluxes.  In addition the ``light
bucket'' method developed for the M87 jet \citep{harris06} to
obtain a rough estimate of the total flux has been also used. 
The energies of all source events in the
evt1 file (i.e. no grade filtering so as to recover events rejected in
evt2 files because of grade migration) from 0.5 to some high energy
such as 11 or 14 keV (determined by inspection) have been summed.  Converting the
resulting values of keV/s to cgs is a very inaccurate process since
it is not possible to know the effective area for each of the photons comprising a
piled event.  Adopting a global effective area of 480 cm$^2$ should
provide a reasonable flux estimate unless these nuclei have an
intrinsic spectrum much harder than that of the knot HST-1 in the M87
jet.  If that were to be the case, we should have used a smaller
effective area, and our fluxes using 480 cm$^2$ would be lower limits.

\subsection{X-ray Spectral Analysis of the stronger nuclei}
\label{sec:spectra}
As already done for the 3C sources observed by \chn\ in Cycles 9 and
12, the X-ray spectral analysis for the nuclei containing 250 or more
counts was performed to estimate their X-ray spectral indices
$\alpha_X$ and the presence or absence of significant intrinsic
absorption N$_H\left(z\right)$.  The spectral analysis was carried out
using the {\sc xspec} version 12.6 software package \citep{arnaud96}.

The spectral data were extracted from a 1\arcsec.5 aperture using the
{\sc ciao} 4.4 routine \texttt{specextract}, thereby automating the creation
of count-weighted response matrices. The background-subtracted spectra
were then filtered in energy between 0.3-7 keV, and binned 
to a minimum of 30 counts per bin to ensure the validity of $\chi^2$ statistics.

Each source was fitted with two multiplicative models: (1) a simple
redshifted powerlaw with Galactic and intrinsic photoelectric
absorption components\\
(\texttt{phabs}$\times$\texttt{zphabs}$\times$\texttt{zpowerlaw} in
{\sc xspec} syntax), \\ and (2) the same model with an additional pileup
component,\\
(\texttt{pileup}$\times$\texttt{phabs}$\times$\texttt{zphabs}$\times$\texttt{zpowerlaw}),
using the {\sc xspec} implementation of the \chn\ pileup
model described by Davis (2001).
The grade migration parameter was set to unity.
For the two piled sources (3C\,327.1 and 3C\,411), the values of
$\alpha_x$ for the jdpileup model are significantly larger (and are to
be preferred) than the values from the standard fits.

Prior to fitting, the Milky Way hydrogen column density
\citep{kalberla05} and the source redshift were fixed to the values
reported in Table~\ref{tab:main}.  The two main variable parameters,
namely the intrinsic absorption N$_H\left(z\right)$ and X-ray spectral
index $\alpha_X$ were allowed to vary in a first pass fit, but
subsequently stepped through a range of possible physical values to
explore the parameter space, determine 90\% confidence intervals, and
quantify the degree to which N$_H\left(z\right)$ and $\alpha_X$ are
degenerate.  Monte Carlo Markov Chains were created to further aid our
understanding of these behaviors.  It is worth noting that those
sources with inverted best-fit spectral indices ($\alpha_X < 0$) can
result from Compton Thick models.  Results are reported in
Table~\ref{tab:spec}.

\begin{table}
\begin{center}
\caption{Spectral Analysis of Bright Nuclei}
\label{tab:spec}
\begin{tabular}{lccccl}
\hline    
          &          &   $N_H$ ($z$)                    &            &              \\ 
Source    &         $\alpha_X$                &   $\left( \times 10^{22} ~\mathrm{cm}^{-2} \right)$ & $\chi^2$/dof \\
  (1)     &          (2)                    &    (3)                           & (4)          \\
\hline 
\noalign{\smallskip}
3C\,67     &      $[0.32^{+0.27}_{-0.25}]$   &      $0.26^{+0.13}_{-0.11}$       &  8.74/15    \\
           &      $[0.51^{+0.31}_{-0.29}]$   &      $0.32^{+0.14}_{-0.12}$       &  8.62/15    \\
3C\,327.1  &      $0.25^{+0.13}_{-0.09}$     &      $0.025^{+0.13}_{\downarrow}$  &  42.7/38    \\                
           &      $0.81^{+0.18}_{-0.17}$     &      $0.16^{+0.15}_{-0.14}$       &  41.5/38    \\
3C\,411    &      $0.11^{+0.06}_{-0.06}$     &      $[0.00^{+0.05}_{\downarrow}]$ &  55.3/60    \\           
           &      $1.01^{+0.16}_{-0.16}$     &      $0.18^{+0.12}_{-0.11}$       &  47.2/60    \\      
3C\,435B    &      $0.96^{+0.28}_{-0.28}$     &      $[3.0^{+1.9}_{-1.7}]$  &  6.38/6     \\     
           &      $1.07^{+0.30}_{-0.30}$     &      $[3.1^{+1.9}_{-1.8}]$  &  6.33/6     \\                    
\noalign{\smallskip}
\hline
\end{tabular}\\
\end{center}
{ These 4 nuclei have 250 or more counts and are thus suitable for spectral analysis.}
The first { row} for each source is the result for a simple power law
fit with galactic and intrinsic absorption.  The second { row} is for
the same model with the addition of `jdpileup'.
For these fits, the corresponding range of best-fit $N_H$ as the spectral index $\alpha_X$ is stepped through values of 0.0 to 2.0 are shown
in square brackets.
Column (1): source name.
Column (2): X-ray spectral index $\alpha_X$.
Column (3): intrinsic absorbing column density in units of 10$^{22}$ cm$^{-2}$.
Column (4): $\chi^2$ / degrees of freedom.
\end{table}

\subsection{Photometric estimates of the intrinsic absorption}
\label{sec:absorption}
Due to the relatively short exposure times of the \chn\ snapshot
survey, it is not often possible to recover the parameters of interest ($\alpha_X$
and N$_H\left(z\right)$) from the spectral fits but it is possible to
derive a range of intrinsic N$_H\left(z\right)$ column densities
corresponding to some chosen range in $\alpha_X$ by using simulated
spectra.  Thus, as already investigated in Paper I and II, 
a photometric analysis was performed to estimate the
intrinsic absorption N$_H\left(z\right)$.  This study
is based on the values of the hardness ratios $HR$ (see
\S~\ref{sec:general}) derived from the photometric analysis of the
nuclear X-ray fluxes.

The observed nuclear fluxes have been used to determine the hardness
ratios $HR$ according to the simple relation: $(H-M)/(H+M)$, where $H$
and $M$ are the X-ray fluxes in the hard and the medium bands,
respectively. The uncertainties on the observed values of $HR$, have
been derived from the X-ray flux errors. We did not use the soft X-ray
band because it is the band most affected by absorption, often leading
to a low number of soft counts.

\begin{figure}[!b]
\includegraphics[scale=0.32,origin=c,angle=0]{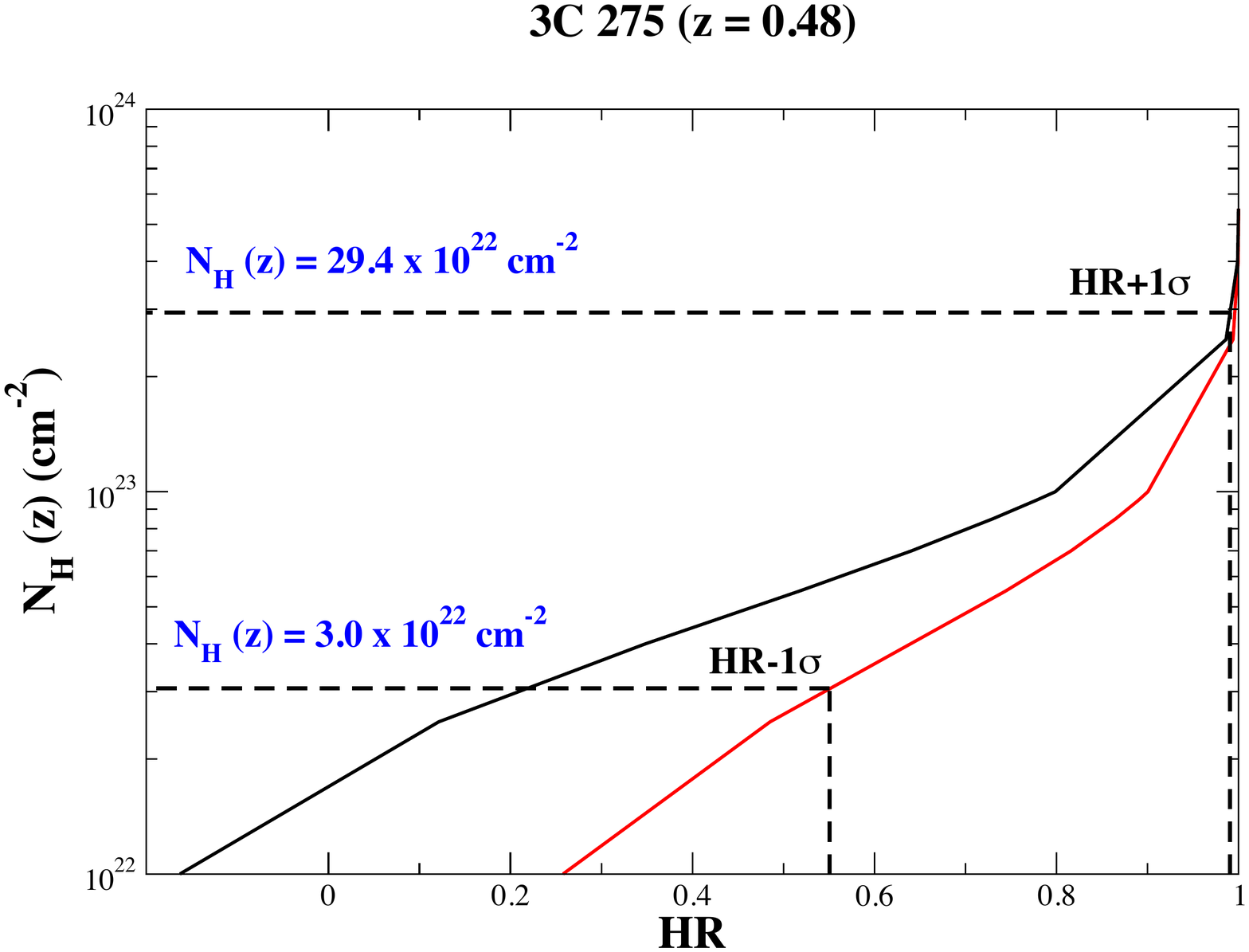}
\caption{The relation between $HR$ and the intrinsic N$_H$ column density
resulting from the simulated spectra at redshift 0.48 as for 3C\,275
and computed for the case of $\alpha_X$ = 0.5 (black solid line) and 1.5 (red solid line).}
\label{fig:nh}
\end{figure}

Most nuclei of radio galaxies show X-ray spectra well described by a
simple power law model with $\alpha_X$ values ranging between 0.5 and
1.5 or occasionally larger \citep[e.g.][]{hardcastle09,worrall09}.  Numerical simulations with {\sc
  xspec} were performed to derive the values of N$_H\left(z\right)$ in
the case of an intrinsically absorbed power-law spectrum with
different values of the spectral index $\alpha_X$ and source redshift
$z$ corresponding to different values of the hardness ratio to derive
the relation between the N$_H\left(z\right)$ and observed $HR$.  We
iterate this procedure for two values of $\alpha_X$ corresponding to
0.5 and 1.5.  In this photometric analysis we adopted a more
restricted energy range of $\alpha_X$ with respect to that used in the
X-ray spectral analysis (see notes to Table~\ref{tab:spec}).  However,
{ this restricted range of $\alpha_X$ values}
is in agreement with previous investigations (i.e., Paper II)
and with the distribution of the spectral index of the low redshift 3C
radio sources \citep[e.g.,][]{hardcastle09}.  In Figure~\ref{fig:nh},
the N$_H$ versus $HR$ curves for 3C\,275 with $\alpha_X$=0.5 and
$\alpha_X$=1.5 are shown.

The N$_H$ estimates corresponding to the
observed $HR$, including 1$\sigma$ error, were calculated 
for the two values of $\alpha_X$ reported above in each source
(see Figure~\ref{fig:nh} for additional details).  
Then, the maximum and the minimum 
values of these N$_H\left(z\right)$ estimates were considered to define the
range where the `real' N$_H\left(z\right)$ value could be, 
corresponding to an estimate of the error on the N$_H\left(z\right)$.
The ranges derived for each $HR$ value are reported in Table~\ref{tab:fluxes}.

It is worth noting that if a generic source is Compton thick (i.e.,
N$_H\left(z\right)>$  10$^{24}$ cm$^{-2}$) or if its X-ray spectrum is inverted
(i.e. $\alpha_X<$ 0) the hardness ratios cannot provide a reliable
estimate of absorption.
{ For a more detailed discussion of the photometric
absorption analysis see Section 3.2 of Paper 2.}

{ Finally, the results obtained from the $HR$ study were compared with
those derived from the X-ray spectral analysis (see
Table~\ref{tab:fluxes} in comparison with Table~\ref{tab:spec}).  
Satisfactory agreement between the two methods was also obtained for the sources in
Paper 2 which were bright enough to warrant spectral analysis.  No comparison is possible
for the two piled sources, 3C327.1 and 3C411.  The single case of disagreement occurred
for the CSS source 3C67, probably}
suggesting a more complex X-ray spectrum.

\section{Results}
\label{sec:results}

\subsection{General}
\label{sec:general}

X-ray emission was detected for all the nuclei in the sample except
for 3C\,435A (see \S~\ref{sec:sources} for more details).  All the
X-ray images are presented in Appendix~\ref{sec:appendixB}.

In addition, as performed for the previous subset of 3C sources
observed during the \chn\ Cycle 9 and 12, the net number of counts
within circular regions of radii 2\arcsec\ and 10\arcsec\ was
measured, both centered on the nucleus of each source.  In
Table~\ref{tab:fluxes} the r=2\arcsec\ result together with the ratio
of r2/r10 are listed, a diagnostic for the presence of extended
emission around the nucleus. The only exceptions are 3C\,19 and
3C\,320 that lie in X-ray detected galaxy clusters (as shown in the
figures of Appendix~\ref{sec:appendixB}) and 3C\,435A
for which the X-ray core is undetected.  This ratio should be close
to unity for an unresolved (i.e., point-like) sources since the
on-axis encircled energy for r=2\arcsec\ is $\approx$0.97, so we expect
only a small increase between r=2\arcsec\ and r=10\arcsec\ for an
unresolved source.

{ Amongst the 19 sources listed in Table~\ref{tab:fluxes}, there are 10 for which the
value of 'Extent Ratio' plus its error is $<$0.9, i.e. indicating the
presence of extra-nuclear emission.  For 6 of these 10, we have
reported features aligned with radio components in Table~\ref{tab:jets}.  Of the
four remaining, 3C\,320 is an obvious cluster of galaxies 
(see Figure~\ref{fig:3c320app} in Appendix~\ref{sec:appendixB}).
For the remaining three (3C\,244.1, 3C\,274.1, and 3C\,306.1) we have
verified the extended emission by way of heavily smoothed images.
There does not appear to be any commonality in the morphologies of the
extended features except that none of the 3 display a smooth,
circularly symmetric emission.  Rather there are irregular clumps of
emission, often, but not always along the radio source axis.}

{ In the current sample of 19 X-ray detected sources all are classified
as FR\,II radio galaxies except for 3C\, 327.1 (uncertain classification)
and the CSS source 3C\,67. }
For 3C\,16, 3C\,19, 3C\,268.2, 3C\,313 we detect hotspots in the
\chn\ images, with confidence levels between 1.4$\sigma$ and
3.2$\sigma$.  In addition, in the cases of 3C\,187 and 3C\,313
extended X-ray emission, of still uncertain nature, is detected,
arising from the regions coincident with, and interior to, their radio
lobes (see Figure~\ref{fig:3c187sm15} and Figure~\ref{fig:3c313sm11},
and \S~\ref{sec:sources} for more details). This extended emission
resembles that associated with radio lobes detected in the X-ray
observations of high redshift radio galaxies \citep[e.g., ][and
  reference therein]{carilli03,smail09,blundell11}.  Finally, X-ray
emission cospatial with radio knots in the jets of 3C\,341
and 3C\,327.1, the last resembling the structure of
3C\,17 \citep{massaro09}, { was also detected}.  Fluxes for jets, hotspots and lobe
structures found in the 3C sample are reported in
Table~\ref{tab:jets}, where the confidence level of each detection
evaluated adopting a Poisson distribution are also provided.

\subsection{Source details}
\label{sec:sources}

\noindent
\underline{{ 3C\,16}} is an FR\,II radio galaxy optically classified
as HEG.  The SW radio lobe is about 16 times the flux density of the
NE lobe at 8 GHz and the brightness ratio is between 30 and 40 for
arcsec sized beams.  From the data available to us, there is no sign
of a radio nucleus nor an X-ray component that would indicate the
position of the host galaxy or quasar.  The original optical
identification \citep{riley80} was the brighter of two faint galaxies.
Since we see no obvious other candidates on an HST image, we have
labelled regions and assigned results assuming that this identification
is correct.  This source is an example of a double-double restarting jet as suggested by 
Schoenmakers et al. (2000) and Gilbert et al. (2004).  On the heavily
smoothed X-ray map, soft X-ray emission was detected roughly aligned
with the two linear segments of excess radio brightness within the SW
lobe and with the brightest part of the southern hotspot (see
Figure~\ref{fig:3c16app} in Appendix~\ref{sec:appendixB}).  The
alignment between X-ray and radio emissions is not accurate, and is
difficult to evaluate quantitatively since the maps have not been
registered.

\noindent
\underline{{ 3C\,19}} is an FR\,II radio galaxy with low excitation
emission lines in its optical spectrum.  The size of the extended
X-ray emission around this source indicates that 3C\,19 lies at the
center of a galaxy cluster.  At this position, we find cluster \#10432
in the list compiled by Wen et al. (2012), based on the Sloan Digital Sky
Survey.  The presence of the galaxy cluster was originally mentioned
by Spinrad et al. (1985).  Within a circle centered on the source with
radius 40\arcsec\ ($\sim$235 kpc), we measured 456$\pm$24 net counts
in the 0.5 - 7 keV energy range.  With the current radio maps
available to us, it was not possible to locate a radio nucleus;
moreover the X-ray core does not appear to be point-like. Thus it was
not possible to register the X-ray image to the radio map.  Enhanced
X-ray brightness associated with the northern hotspot and perhaps also
at the southern hotspot has been found, while the axis of the radio
and of the X-ray emission is directed almost perpendicular to the
optical emission seen by HST \citep{dekoff96}.

\noindent
\underline{{ 3C\,42}} is an FR\,II radio galaxy optically classified
as HEG.  The X-ray nucleus was clearly detected in the \chn\ snapshot observation.

\noindent
\underline{{ 3C\,46}} is an FR\,II - HEG radio galaxy.  There was some
uncertainty locating the nucleus for the host galaxy of 3C\,46.
Examining the low resolution 1.5 GHz radio map, one would confidently
identify the nucleus as the weak source between the two lobes.
However, at 8.4 GHz with an 0.3\arcsec\ beam, there is no detectable
emission at this location.  Instead, there is an unresolved source
(1.9 mJy) 6.7\arcsec\ to the south-west, just inside the western lobe
and very close to the line joining the southern hotspot to the weak
1.5 GHz source between the lobes.  The X-ray core aligns with this 8.4
GHz source.  Although it is possible to posit very heavy absorption so
as to render the nucleus unobservable in the \chn\ bandpass, it is
puzzling that the nucleus would be detected at 1.5 GHz but not at 8.4
GHz.  Of the 51 counts associated with the 8.4 GHz source, only 6 are
below 2 keV, with 9 in the 2-4keV band and 36 between 4 and 7 keV (no
counts above 7 keV).  For a power law distribution, both the radio and
the X-ray spectra would be inverted (i.e., $\alpha~<$~0).  According
to Spinrad et al. (1985) 3C\,46 belongs to a galaxy cluster but no
extended X-ray emission is detected in the \chn\ observations.

\noindent
\underline{{ 3C\,67}} is the only CSS radio source in this sample, and
is optically classified as a Broad Line Object (BLO). The bright X-ray
nucleus is the only detection.  The optical emission revealed by HST
is extended in the same direction as the radio axis \citep{dekoff96}.

\begin{figure}
\includegraphics[scale=0.45,origin=c,angle=0]{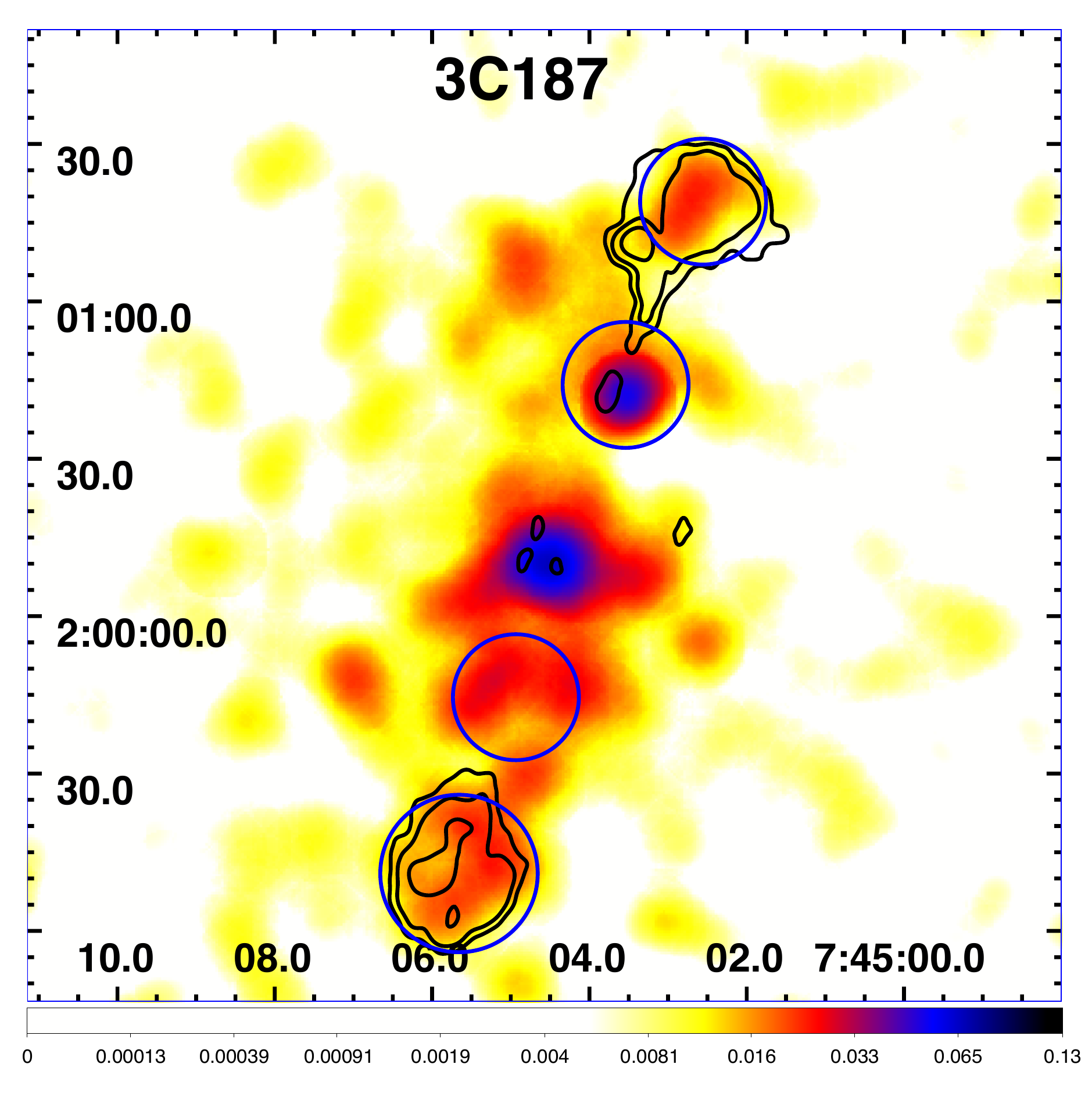}
\caption{3C\,187 (0.5-7keV) smoothed with a Gaussian of FWHM=11$^{\prime\prime}$.
The color bar gives the brightness in units of counts/pixel and the
pixel size is 0.492$^{\prime\prime}$.  The four regions listed in
Table~\ref{tab:jets} are shown in blue.  The contours come from an L band radio map
with a clean beam of 3$^{\prime\prime}$ and start at 1 mJy/beam,
increasing by factors of 4.}
\label{fig:3c187sm15}
\end{figure}

\noindent
\underline{{ 3C\,187}} is a typical FR\,II radio galaxy.  Two
unresolved radio sources lie where the radio
nucleus is expected to be.  They are separated by 2.0\arcsec\ in PA=-21$^{\circ}$;
i.e. along the principle axis of the radio source.  Comparing flux
densities at 1.4 GHz and 8.4 GHz, it is clear that the southern source
has an inverted spectrum whereas the northern source has a normal
spectrum.  The southern source is located at RA=07h 45m 04.46s,
DEC=2$^{\circ}$ 00$^{\prime}$ 08.7\arcsec\ with flux densities
S$_{1.4}$=0.9mJy and S$_{8.4}$=3.3mJy.
The northern radio source has
S$_{1.4}$=0.9mJy and S$_{8.4}$=0.4mJy.  The nucleus is not well
defined in the X-rays; there are only two counts above 2keV coincident
with the southern source.  In the heavily smoothed X-ray map (see
Figure~\ref{fig:3c187sm15}) significant X-ray emission from
both the northern and the southern lobes appears; 431$\pm$27 net
counts were found between 0.5 and 7 keV, mostly below 2 keV within a rectangular
region 170\arcsec\ x 75\arcsec\ (PA=-21$^{\circ}$) with two background
regions of the same size deployed on either side.

\noindent
\underline{{ 3C\,244.1}} The fact that the soft flux is significantly
larger than the medium flux suggests the presence of a soft excess.
On the other hand, no extended emission arising from its surrounding
galaxy cluster \citep{hill91} is seen in the \chn\ observation.

\noindent
\underline{{ 3C\,268.2}} is an FR\,II radio galaxy with no nuclear
radio emission detectable in the 1.4 GHz map available to us.  An
X-ray source lies between the radio emission arising from the hotspot
regions so is plausibly the X-ray core.  The X-ray nucleus is elongated
in the EW direction. In the HST image the source has strong extended
emission-line regions which could be responsible for its apparent
elongated morphology \citep{dekoff96}.  In addition, a marginal
detection of the X-ray counterpart of the southern hotspot has been
found.
 
\noindent
\underline{{ 3C\,274.1}} is an FR\,II radio galaxy that lies in a
galaxy cluster \citep{yates89}.  At this position, we find cluster
\#95983 in the list compiled by Wen et al. (2012), based on the Sloan
Digital Sky Survey.  However, in the X-rays, only the nucleus (that
appears to be extended) is detected.

\noindent
\underline{{ 3C\,275}} is an FR\,II radio galaxy optically classified
as LEG.  In all the radio maps available to us ranging from 1.4 GHz up
to 14.9 GHz, the radio nucleus is not detected, thus the alignment
between X-ray and radio emissions is not precise, since the maps have
not been registered.  The extended X-ray emission arising from the
core of 3C\,275 was clearly detected.  The original optical
identification \citep{kristian74} reports the galaxy is in a distant
cluster but we have no direct evidence for that.

\noindent
\underline{{ 3C\,306.1}} is a radio galaxy optically classified as HEG
and radio classified as FR\,II.  In the \chn\ observation,
the X-ray nucleus, which is not point-like, was seen. 
 3C\, 306.1 is also a member of a galaxy cluster \citep{yates89} but its X-ray thermal
emission was not seen in the 3C snapshot survey.
 
\begin{figure}
\includegraphics[scale=0.45,origin=c,angle=0]{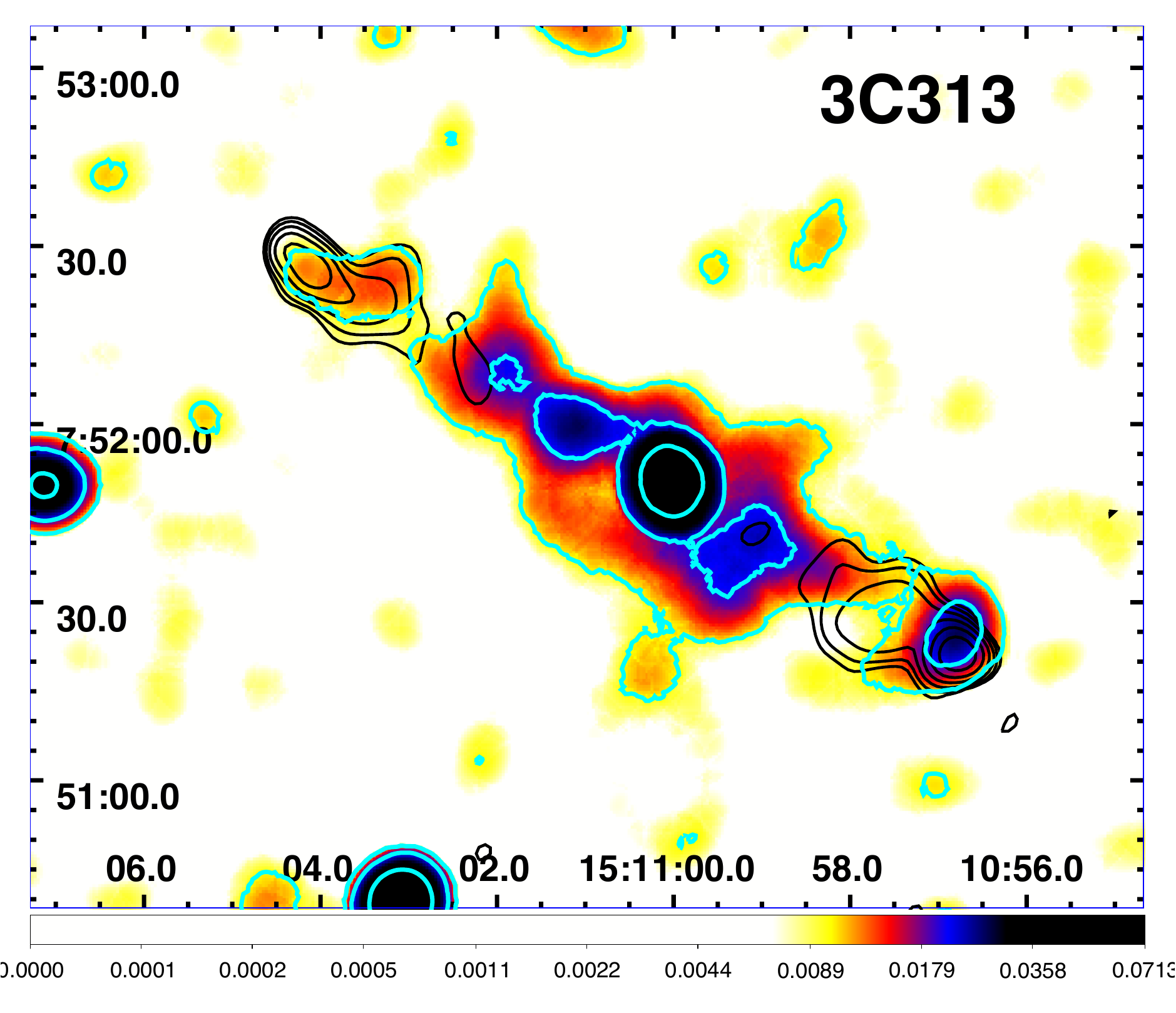}
\caption{An X-ray image of 3C\,313 for the energy band 0.5-7 keV.  The pixel
size is 0.492\arcsec\ and the smoothing function is a Gaussian of
FWHM=11.0\arcsec.  X-ray contours (white or cyan) start at
0.01 counts/pix and increase by factors of two. The radio contours
(black) come from a 1.4 GHz map constructed from archival VLA data and
start at 15 mJy/beam, increasing by factors of two.  The clean beam is
5\arcsec.}
\label{fig:3c313sm11}
\end{figure}

\noindent
\underline{{ 3C\,313}} is an FR\,II radio galaxy optically classified
as an HEG. The galaxy nucleus is surrounded by patches of optical
emission in the HST images \citep{dekoff96}.  Although no clear
signatures of X-ray cluster emission was found, 3C\,313 is a member of
a galaxy cluster \citep{hill91,stanford02}.  At this position, we find
cluster \#107171 in the list compiled by Wen et al. (2012), based on the
Sloan Digital Sky Survey. There is, however enhanced X-ray brightness
along the entire principle axis of the radio source; i.e. even close
to the nucleus where no radio emission is seen with the data available
to us (VLA, 1.4 GHz and 8.4 GHz) (see Figure~\ref{fig:3c313sm11}).  In
addition to this extended emission which could be either thermal
(e.g., extended optical emission line region) or non-thermal (e.g.,
inverse Compton emission from radio emitting electrons scattering CMB
photons), 4 counts were detected near the tip of the northern hotspot
and 9 counts are aligned with the southern hotspot.

\noindent
\underline{{ 3C\,320}} is a classical FR\,II radio galaxy located in
the center of a cluster of galaxies \citep{spinrad85}.  At this
position, we find cluster \#115229 in the list compiled by Wen et al. (2012),
based on the Sloan Digital Sky Survey. 3C\,320 shares many attributes with
Cygnus A.  The eastern hotspot/lobe has engendered an obvious cavity
in the cluster gas. In the \chn\ observations of 12ksec, 1046$\pm$37
net counts were found in the 0.5 - 7 keV energy range within a
circular region of 1\arcmin\ radius (286 kpc).
 
\noindent
\underline{{ 3C\,327.1}} is optically classified as HEG hosted in a
normal elliptical galaxy \citep{dekoff96}.  This radio source has a
curved jet on the SE side, and only a low brightness lobe to the
NW. Its radio structure \citep{morganti99} is remarkably similar to
that of the low redshift radio source 3C\,17 \citep{massaro09}.  The
X-ray emission is extended around the nucleus and a trace of the
southern jet can be seen in the X-ray image (see Figure in
Appendix~\ref{sec:appendixB}).

\noindent
\underline{{ 3C\,341}} is a classical FR\,II radio galaxy optically classified as HEG. 
High resolution radio maps at 8 GHz show a slightly curved line of enhanced brightness
over the first half of the SW lobe.  At $\approx~8^{\prime\prime}$ from the nucleus
there is a knot $\approx$ 4 times brighter than adjoining areas.  At the position of
this knot in the jet there are 2.5$\pm$1.8 net X-ray counts.
 
\noindent
\underline{{ 3C\,411}} is an FR\,II radio galaxy with high excitation
emission lines in its optical spectrum.  The X-ray core is the
brightest detected in the current sample with $\approx$~0.75
counts/frame and is thus affected by substantial pileup. 
The ratio of evt1 counts to evt2 counts is 1.26.

\noindent
\underline{{ 3C\,434}} is an FR\,II radio galaxy hosted in an
elliptical galaxy, whose optical emission is slightly elongated in the
southeast direction \citep{dekoff96}.  This radio galaxy lies close to
a galaxy group according to Yates et al. (1989), however, the X-ray
snapshot observation only revealed nuclear emission.

\noindent
\underline{{ 3C\,435}} is comprised of two double-lobed radio
galaxies. 3C435A is an FR\,II radio galaxy at redshift 0.471
\citep{spinrad85} and 3C435B is another FR\,II radio source lying
10\arcsec\ away but at z=0.865 \citep{mccarthy89}. The X-ray core of
3C\,435A is undetected in the \chn\ snapshot observation, while X-ray
emission was detected from the nuclear region of 3C\,435B.  It is
worth noting that a blue stellar object (A type star) lies in the
direction of 3C\,435B \citep{mccarthy89}, however given the
significant X-ray emission above $\sim$2 keV, we suspect the
\chn\ detection arises from the host galaxy of 3C\,435B.

\section{Summary}
\label{sec:summary}
We have presented the X-ray analyses of 19 3C radio sources with
redshift between 0.3 and 0.5. Since proprietary rights were waived,
X-ray data based on a radio flux limited sample (almost all extragalactic 3C
sources with $z<$0.5) are now available to the community.  The main
objectives of this 3C snapshot survey are: to detect X-ray emission
from jets and hotspots, and to study the nuclear emission of their
host galaxies.  In the present work, the basic source parameters for
the newly acquired data are presented, while in future works, the
resulting datasets will be used to test several issues such as the
Fanaroff-Riley dichotomy and differences in the nature of nuclear
absorption in FR\,I and FR\,II sources.

Fluxmaps for all the X-ray observations were constructed and
we provide photometric results for the nuclei and other radio structures
(i.e., jet knots, hotspots, lobes). 
{ Using hardness ratios, we have estimated the column density for
intrinsic absorption for 16 nuclei (2 nuclei are affected by pileup
and one has too few counts).}
In addition, for the { 4 brighest} nuclei, X-ray spectral analysis was performed, comparing the
results on the intrinsic absorption with those derived from the
hardness ratio analysis (see \S~\ref{sec:absorption}).

X-ray emission was detected for all the nuclei except
3C\,435A.  A sizable fraction ($\sim$60\%) of them show evidence for
significant intrinsic absorption (\S~\ref{sec:spectra} and
\S~\ref{sec:absorption}).  Amongst these 3C \chn\ observations, we
found X-ray emission arising from one compact steep spectrum radio
sources: 3C\,67 and X-ray emission from two galaxy clusters
surrounding 3C\,19 and 3C\,320.  In addition, we find X-ray emission
from six radio hotspots (in the four radio galaxies 3C\,16, 3C\,19,
3C\,268.2 and 3C\,313) and several knots in the radio jets of
3C\,327.1 and 3C\,341.  Finally, in the FR\,II radio galaxies 3C\,187
and 3C\,313 extended X-ray emission, of still unknown nature, was
discovered, arising from regions along the principle radio axis (see
\S~\ref{sec:sources} for more details) .

\acknowledgments 
We thank the anonymous referee for useful comments that led to improvements in the paper.
We are grateful to R. Morganti, M. Hardcastle and C. C. Cheung for providing
several radio maps of the 3C sources and for helpful discussions.
This research has made use of NASA's Astrophysics Data System.
TOPCAT\footnote{\underline{http://www.star.bris.ac.uk/$\sim$mbt/topcat/}}
\citep{taylor2005} and SAOImage DS9 were used extensively in this work
for the preparation and manipulation of the tabular data and the
images.  SAOImage DS9 was developed by the Smithsonian Astrophysical
Observatory.  The NASA/IPAC Extragalactic Database (NED) is provided
by the Jet Propulsion Laboratory, California Institute of Technology,
under contract with the National Aeronautics and Space Administration.
Several radio maps were downloaded from the NVAS (NRAO VLA Archive
Survey) and from the DRAGN
webpage\footnote{http://www.jb.man.ac.uk/atlas/}.  The National Radio
Astronomy Observatory is operated by Associated Universities, Inc.,
under contract with the National Science Foundation.   The work at SAO is
supported by NASA-GRANTS GO1-12125A and GO2-13115X.  This work was
also supported by contributions of European Union, Valle DÕAosta
Region and the Italian Minister for Work and Welfare. AB acknowledges
support by the research group FOR 1254 funded by the Deutsche
Forschungsgemeinschaft: `Magnetization of interstellar and
intergalactic media: the prospect of low-frequency radio observationsÕ.  
G.R.T acknowledges support by the European Community's Seventh
Framework Programme (/FP7/2007- 2013/) under grant agreement
No. 229517.\\

{Facilities:} \facility{VLA}, \facility{CXO (ACIS)}

\appendix
\section{A. Chandra observations of the 3CR sample}
\label{sec:appendixA}
In Table~\ref{tab:3CR} we list the summary of the \chn\ observations for the whole 3CR sample.
We report the 3CR name together with an alternate name for the most observed sources,
the radio classification, the coordinates, the redshift (if known), the number of \chn\ observations
as well as the reference for the first \chn\ image and some X-rays notes.
In particular, we marked 3CR sources that lie in a galaxy cluster (cl) 
and if the cluster is also detected in the X-rays (xcl), 
and/or if a jet knot, hotspot, lobe is also detected in the X-rays (k,h,l, respectively) as reported in the literature.
The radio classification is based on the Spinrad et al. (1985) version of the 3CR catalog,
however we distinguished between FR\,I and FR\,II radio galaxies \citep[e.g.,][]{fanaroff74}.
Uncertain classification are labeled as ``UND'' while uncertain or unknown redshift estimates are marked with ``?''. 
If the 3CR source is not observed by \chn\ we simply reported a dashed line ``-''; when no reference are listed 
but the number of \chn\ observations is grater than 1 implies that the X-ray observations have not been performed or not yet been published. 
The absence of one or more codes in the X-ray notes does not necessarily mean there is no detection of extended emission. 

References for the radio positions used in Table~\ref{tab:3CR} are:
Becker et al. (1995), Best et al. (1999), Condon et al. (1998), Djorgovski et al. (1988), Douglas et al. (1996),
Evans et al. (2010a), Fernini et al. (1997), Fey et al. (2004), Fomalont et al. (2003), Hiltner \& Roser (1991),
Jackson et al. (2007), Johnston et al. (1995), Paris et al. (2012), Petrov et al. (2005), 
Taylor et al. (2001), White et al. (1997).

\begin{table*}
\tiny
\label{tab:3CR}
\begin{tabular}{rrrlllcll}
\hline
3CR  & Other & Class & R.A.    & Dec.    & z & \chn\  & Reference & X-ray \\
name &  name &       & (J2000) & (J2000) &  & obs. &           & notes \\
\hline 
\noalign{\smallskip}
  2     & -        & QSR   & 00 06 22.588 & -00 04 24.69 & 1.038  & 1   & Shemmer et al. (2006)      & -           \\     
  6.1   & -        & FR II & 00 16 31.070 & +79 16 49.99 & 0.84   & 2   & Hardcastle et al. (2004)   & h           \\
  9     & -        & QSR   & 00 20 25.219 & +15 40 54.59 & 2.009  & 1   & Fabian et al. (2003)       & k           \\
 11.1   & -        & UND   & 00 29 44.808 & +63 58 42.81 & ?      & -   & -                          & -           \\
 13     & -        & FR II & 00 34 14.556 & +39 24 16.65 & 1.351  & 1   & Wilkes et al. (2013)       & -           \\ 
 14     & -        & QSR   & 00 36 06.447 & +18 37 59.23 & 1.469  & 1   & Wilkes et al. (2013)       & -           \\ 
 14.1   & -        & UND   & 00 36 27.12  & +59 46 30.00 & ?      & -   & -                          & -           \\
 15     & -        & FR I  & 00 37 04.114 & -01 09 08.46 & 0.074  & 1   & Kataoka et al. (2003a)     & k,l         \\
 16     & -        & FR II & 00 37 45.40  & +13 20 09.2  & 0.405  & 1   & this work                  & h,l         \\
 17     & -        & QSR   & 00 38 20.528 & -02 07 40.49 & 0.2197 & 1   & Massaro et al. (2009a)     & k           \\
\hline 
\noalign{\smallskip}
 18     & -        & FR II & 00 40 50.553 & +10 03 26.78 & 0.188  & 1   & Massaro et al. (2010)      & -           \\ 
 19     & -        & FR II & 00 40 55.044 & +33 10 08.02 & 0.482  & 1   & this work                  & cl,h        \\
 20     & -        & FR II & 00 43 09.177 & +52 03 36.05 & 0.174  & 1   & Massaro et al. (2010)      & -           \\
 21.1   & -        & UND   & 00 44 41.299 & +66 18 42.31 & ?      & -   & -                          & -           \\
 22     & -        & FR II & 00 50 56.301 & +51 12 03.01 & 0.936  & 1   & -                          & -           \\
 27     & -        & UNC   & 00 56 01.058 & +68 22 30.42 & 0.184  & -   & -                          & -           \\
 28     & -        & FR I  & 00 55 50.58  & +26 24 36.6  & 0.1952 & 5   & Donato et al. (2004)       & xcl         \\
 29     & -        & FR I  & 00 57 34.895 & -01 23 27.37 & 0.0448 & 1   & Massaro et al. (2012)      & cl,k        \\
 31     &NGC\,383  & FR I  & 01 07 24.959 & +32 24 45.21 & 0.017  & 1   & Hardcastle et al. (2002a)  & cl,k        \\
 33     & -        & FR II & 01 08 52.858 & +13 20 14.20 & 0.0059 & 1   & Evans et al. (2010b)       & h           \\
\hline 
\noalign{\smallskip}
 33.1   & -        & FR II & 01 09 44.237 & +73 11 57.10 & 0.181  & 2   & Massaro et al. (2010)      & -           \\
 33.2   & -        & UND   & 01 08 34.060 & +69 22 33.80 & ?      & -   & -                          & -           \\
 34     & -        & FR II & 01 10 18.666 & +31 47 20.47 & 0.69   & -   & -                          & -           \\
 35     & -        & FR II & 01 12 02.236 & +49 28 35.03 & 0.067  & 1   & -                          & -           \\
 36     & -        & QSR   & 01 17 59.480 & +45 36 21.75 & 1.301  & -   & -                          & -           \\
 40     &ARP\,308  & FR II & 01 25 53.50  & -01 21 54.00 & 0.018  & 2   & Bogdan et al. (2011)       & xcl         \\
 41     & -        & FR II & 01 26 44.389 & +33 13 11.21 & 0.795  & -   & -                          & -           \\ 
 42     & -        & FR II & 01 28 30.12  & +29 03 00.8  & 0.395  & 1   & this work                  & -           \\
 43     & -        & QSR   & 01 29 59.809 & +23 38 20.28 & 1.459  & 1   & Wilkes et al. (2013)       & -           \\ 
 44     & -        & FR II & 01 31 21.765 & +06 23 40.82 & 0.66   & -   & -                          & cl          \\
\hline 
\noalign{\smallskip}
 46     & -        & FR II & 01 35 28.48  & +37 54 05.2  & 0.4373 & 1   & this work                  & cl          \\ 
 47     & -        & QSR   & 01 36 24.411 & +20 57 27.44 & 0.425  & 1   & Hardcastle et al. (2004)   & h           \\
 48     & -        & QSR   & 01 37 41.299 & +33 09 35.13 & 0.367  & 1   & Siemiginowska et al. (2008)& -           \\
 49     & -        & FR II & 01 41 09.160 & +13 53 28.05 & 0.621  & 1   & -                          & -           \\
 52     & -        & FR II & 01 48 28.909 & +53 32 28.04 & 0.2854 & 1   & Massaro et al. (2010)      & cl,h        \\
 54     & -        & FR II & 01 55 30.162 & +43 45 55.43 & 0.8274 & -   & -                          & -           \\
 55     & -        & FR II & 01 57 10.510 & +28 51 37.56 & 0.7348 & -   & -                          & -           \\
 61.1   & -        & FR II & 02 22 35.571 & +86 19 06.38 & 0.1878 & 1   & Massaro et al. (2010)      & h           \\  
 63     & -        & FR II & 02 20 54.316 & -01 56 50.72 & 0.175  & 1   & Massaro et al. (2012)      & -           \\
 65     & -        & FR II & 02 23 43.191 & +40 00 52.45 & 1.176  & 1   & Wilkes et al. (2013)       & -           \\ 
\hline 
\noalign{\smallskip}
 66A    & -        & BL    & 02 22 39.611 & +43 02 07.79 & ?      & 2   & -                          & -           \\
 66B    &UGC\,1841 & FR I  & 02 23 11.411 & +42 59 31.38 & 0.0213 & 1   & Hardcastle et al. (2001)   & cl,k        \\ 
 67     & -        & FR II & 02 24 12.29  & +27 50 11.5  & 0.310  & 1   & this work                  & -           \\ 
 68.1   & -        & QSR   & 02 32 28.872 & +34 23 46.79 & 1.238  & 1   & Wilkes et al. (2013)       & -           \\  
 68.2   & -        & FR II & 02 34 23.856 & +31 34 17.46 & 1.575  & 1   & Wilkes et al. (2013)       & -           \\
 69     & -        & FR II & 02 38 02.355 & +59 11 50.00 & 0.458  & -   & -                          & -           \\
 71     &NGC\,1068 & Sy    & 02 42 40.711 & -00 00 47.81 & 0.0037 & 15  & Brinkman et al. (2002)     & -           \\
 75     &NGC\,1128 & FR I  & 02 57 41.57  & +06 01 28.8  & 0.0231 & 1   & Hudson et al. (2006)       & xcl         \\
 76.1   & -        & FR II & 03 03 15.054 & +16 26 18.83 & 0.0324 & 1   & Massaro et al. (2010)      & -           \\
 78     &NGC\,1218 & FR I  & 03 08 26.223 & +04 06 39.30 & 0.0286 & 2   & Harwood \& Hardcastle (2012) &  -        \\
\hline 
\noalign{\smallskip}
 79     & -        & FR II & 03 10 00.090 & +17 05 58.52 & 0.2559 & 1   & Massaro et al. (2012)      & cl          \\
 83.1   &NGC\,1265 & FR I  & 03 18 15.664 & +41 51 27.88 & 0.0251 & 1   & Sun et al. (2005)          & xcl,k       \\
 84     &NGC\,1275 & FR I  & 03 19 48.160 & +41 30 42.10 & 0.0175 & 27  & Fabian et al. (2002)       & xcl         \\ 
 86     & -        & FR II & 03 27 19.357 & +55 20 28.12 & ?      & -   & -                          & -           \\    
 88     &UGC\,2478 & FR I  & 03 27 54.194 & +02 33 41.98 & 0.0302 & 4   & Sun (2009)                 & xcl         \\
 89     & -        & FR I  & 03 34 15.574 & -01 10 56.09 & 0.1386 & 1   & Massaro et al. (2012)      & xcl         \\
 91     & -        & UND   & 03 37 43.353 & +50 45 52.83 & ?      & -   & -                          & -           \\
 93     & -        & QSR   & 03 43 30.008 & +04 57 48.58 & 0.3571 & -   & -                          & -           \\ 
 93.1   & -        & UNC   & 03 48 46.934 & +33 53 15.28 & 0.2430 & 1   & Massaro et al. (2012)      & cl          \\
 98     & -        & FR II & 03 58 54.442 & +10 26 03.03 & 0.0305 & 1   & Hodges-Kluck et al. (2010) & -           \\
\noalign{\smallskip}
\hline
\end{tabular}
\end{table*} 

\begin{table*}
\tiny
\label{tab:3CR}
\begin{tabular}{rrrlllcll}
\hline
3CR  & Other & Class & R.A.    & Dec.    & z & \chn\  & Reference & X-ray \\
name &  name &       & (J2000) & (J2000) &  & obs. &           & notes \\
\hline 
\noalign{\smallskip}
 99     & -        & Sy    & 04 01 07.628 & +00 36 32.91 & 0.426  & 1   & -                          & cl          \\
103     & -        & FR II & 04 08 03.10  & +43 00 32.70 & 0.330  & 1   & this work                  & -           \\ 
105     & -        & FR II & 04 07 16.453 & +03 42 25.80 & 0.089  & 1   & Massaro et al. (2010)      & k,h         \\
107     & -        & FR II & 04 12 22.627 & -00 59 32.43 & 0.785  & -   & -                          & -           \\
109     & -        & FR II & 04 13 40.37  & +11 12 13.8  & 0.3056 & 1   & Hardcastle et al. (2004)   & h           \\
111     & -        & FR II & 04 18 21.277 & +38 01 35.80 & 0.0485 & 2   & Hogan et al. (2011)        & k,h         \\
114     & -        & FR II & 04 20 22.170 & +17 53 55.20 & 0.815  & -   & -                          & -           \\
119     & -        & UNC   & 04 32 36.502 & +41 38 28.44 & 1.023  & -   & -                          & -           \\
123     & -        & FR II & 04 37 04.375 & +29 40 13.81 & 0.2177 & 1   & Hardcastle et al. (2001)   & cl,k        \\
124     & -        & UNC   & 04 41 59.108 & +01 21 01.91 & 1.083  & -   & -                          & -           \\
\hline 
\noalign{\smallskip}
125     & -        & UND   & 04 46 17.857 & +39 45 03.03 & ?      & -   & -                          & -           \\
129     &WEIN\,045 & FR I  & 04 49 09.076 & +45 00 39.22 & 0.0208 & 2   & Krawczynski (2002)         & xcl,k       \\
129.1   &WEIN\,051 & FR I  & 04 50 06.67  & +45 03 05.8  & 0.0222 & 1   & Krawczynski (2002)         & -           \\
130     & -        & FR I  & 04 52 52.836 & +52 04 47.09 & 0.1090 & 1   & Massaro et al. (2012)      & -           \\ 
131     & -        & UND   & 04 53 23.328 & +31 29 25.30 & ?      & -   & -                          & -           \\
132     & -        & FR II & 04 56 42.919 & +22 49 23.20 & 0.214  & 1   & Massaro et al. (2010)      & cl          \\
133     & -        & FR II & 05 02 58.472 & +25 16 25.31 & 0.2775 & 1   & Massaro et al. (2010)      & -           \\
134     & -        & UND   & 05 04 42.19  & +38 06 11.4  & ?      & -   & -                          & -           \\
135     & -        & FR II & 05 14 08.367 & +00 56 32.28 & 0.1273 & 1   & Massaro et al. (2010)      & cl          \\
136.1   & -        & FR II & 05 16 03.14  & +24 58 25.5  & 0.064  & 1   & Balmaverde et al. (2012)   & -           \\
\hline 
\noalign{\smallskip}
137     & -        & UND   & 05 19 32.422 & +50 54 32.08 & ?      & -   & -                          & -           \\
138     & -        & QSR   & 05 21 09.886 & +16 38 22.05 & 0.759  & 1   & -                          & -           \\
139.2   & -        & FR II & 05 24 27.11  & +28 12 47.0  & ?      & -   & -                          & -           \\
141     & -        & UND   & 05 26 44.20  & +32 50 23.0  & ?      & -   & -                          & -           \\
142.1   & -        & UNC   & 05 31 29.375 & +06 30 24.99 & 0.406  & -   & -                          & -           \\
147     & -        & QSR   & 05 42 36.137 & +49 51 07.23 & 0.545  & 1   & -                          & -           \\
152     & -        & UND   & 06 04 28.628 & +20 21 21.73 & ?      & -   & -                          & -           \\
153     & -        & FR II & 06 09 32.423 & +48 04 14.64 & 0.2769 & 1   & Massaro et al. (2010)      & cl          \\
154     & -        & QSR   & 06 13 50.139 & +26 04 36.72 & 0.58   & -   & -                          & -           \\
158     & -        & UND   & 06 21 41.063 & +14 32 11.19 & ?      & -   & -                          & -           \\
\hline 
\noalign{\smallskip}
165     & -        & FR II & 06 43 07.400 & +23 19 03.00 & 0.2957 & 1   & Massaro et al. (2010)      & -           \\ 
166     & -        & FR II & 06 45 24.098 & +21 21 51.30 & 0.2449 & 1   & Massaro et al. (2012)      & -           \\
169.1   & -        & FR II & 06 51 15.355 & +45 09 26.22 & 0.633  & -   & -                          & -           \\
171     & -        & FR II & 06 55 14.722 & +54 08 57.46 & 0.2384 & 1   & Massaro et al. (2010)      &  k,h        \\
172     & -        & FR II & 07 02 08.075 & +25 13 46.30 & 0.5191 & 1   & -                          & -           \\
173     & -        & QSR   & 07 02 17.60  & +37 57 19.5  & 1.035  & -   & -                          & -           \\
173.1   & -        & FR II & 07 09 18.084 & +74 49 31.89 & 0.2921 & 1   & Hardcastle et al. (2004)   & cl,h        \\
175     & -        & QSR   & 07 13 02.400 & +11 46 14.65 & 0.77   & 1   & -                          & -           \\
175.1   & -        & FR II & 07 14 04.682 & +14 36 22.00 & 0.92   & 1   & -                          & -           \\
180     & -        & FR II & 07 27 04.880 & -02 04 30.33 & 0.22   & 1   & Massaro et al. (2012)      & -           \\
\hline 
\noalign{\smallskip}
181     & -        & QSR   & 07 28 10.305 & +14 37 36.24 & 1.382  & 1   & Wilkes et al. (2013)       & -           \\
184     & -        & FR II & 07 39 24.471 & +70 23 10.88 & 0.994  & 1   & -                          & xcl         \\
184.1   & -        & FR II & 07 43 01.394 & +80 26 26.09 & 0.1182 & 1   & Massaro et al. (2010)      & cl          \\
186     & -        & QSR   & 07 44 17.452 & +37 53 17.15 & 1.067  & 5   & Siemiginowska et al. (2005)& xcl         \\
187     & -        & FR II & 07 45 04.46  & +02 00 08.70 & 0.350  & 1   & this work                  & l           \\  
190     & -        & QSR   & 08 01 33.559 & +14 14 42.94 & 1.1944 & 1   & Wilkes et al. (2013)       & -           \\
191     &0802+103  & QSR   & 08 04 47.972 & +10 15 23.69 & 1.956  & 2   & Sambruna et al. (2004)     & l           \\
192     & -        & FR II & 08 05 34.996 & +24 09 50.34 & 0.0597 & 1   & Hodges-Kluck et al. (2010) & cl          \\
194     & -        & UNC   & 08 10 03.619 & +42 28 04.31 & 1.184  & -   & -                          & -           \\
196     & -        & QSR   & 08 13 36.033 & +48 13 02.56 & 0.871  & 1   & -                          & -           \\
\hline 
\noalign{\smallskip}
196.1   & -        & FR II & 08 15 28.10  & -03 08 28.00 & 0.198  & 1   & Massaro et al. (2012)      & -           \\
197.1   & -        & FR II & 08 21 33.605 & +47 02 37.40 & 0.1301 & 1   & Massaro et al. (2010)      & cl          \\
198     & -        & FR II & 08 22 31.80  & +05 57 07.90 & 0.0815 & 1   & Massaro et al. (2012)      & cl          \\
200     & -        & FR II & 08 27 25.397 & +29 18 44.90 & 0.458  & 1   & Hardcastle et al. (2004)   & h           \\
204     & -        & QSR   & 08 37 44.956 & +65 13 34.92 & 1.112  & 1   & Wilkes et al. (2013)       & -           \\
205     & -        & QSR   & 08 39 06.459 & +57 54 17.12 & 1.534  & 1   & Wilkes et al. (2013)       & -           \\
207     & -        & QSR   & 08 40 47.588 & +13 12 23.56 & 0.6808 & 2   & Brunetti et al. (2002)     & h           \\
208     & -        & QSR   & 08 53 08.609 & +13 52 54.98 & 1.1115 & 1   & Wilkes et al. (2013)       & -           \\
208.1   & -        & UNC   & 08 54 39.289 & +14 05 52.56 & 1.02   & -   & -                          & -           \\
210     & -        & UNC   & 08 58 09.961 & +27 50 51.57 & 1.169  & 1   & -                          & xcl         \\
\noalign{\smallskip}
\hline
\end{tabular}
\end{table*} 

\begin{table*}
\tiny
\label{tab:3CR}
\begin{tabular}{rrrlllcll}
\hline
3CR  & Other & Class & R.A.    & Dec.    & z & \chn\  & Reference & X-ray \\
name &  name &       & (J2000) & (J2000) &  & obs. &           & notes \\
\hline 
\noalign{\smallskip}
212     & -        & QSR   & 08 58 41.498 & +14 09 43.97 & 1.048  & 1   & Aldcroft et al. (2003)     & h           \\
213.1   & -        & FR I  & 09 01 05.269 & +29 01 46.88 & 0.1937 & 1   & Massaro et al. (2010)      & cl,h        \\
215     & -        & QSR   & 09 06 31.873 & +16 46 11.87 & 0.4121 & 1   & Hardcastle et al. (2004)   & -           \\ 
216     & -        & QSR   & 09 09 33.497 & +42 53 46.48 & 0.6702 & 1   & -                          & -           \\
217     & -        & FR II & 09 08 50.583 & +37 48 19.21 & 0.8975 & -   & -                          & -           \\
219     & -        & FR II & 09 21 08.623 & +45 38 57.39 & 0.1744 & 1   & Comastri et al. (2003)     & cl,k        \\
220.1   & -        & FR II & 09 32 39.646 & +79 06 31.53 & 0.61   & 1   & Worrall et al. (2001)      & xcl         \\
220.2   & -        & QSR   & 09 30 33.473 & +36 01 24.17 & 1.1577 & -   & -                          & -           \\ 
220.3   & -        & FR II & 09 39 22.52  & +83 15 24.5  & 0.68   & 1   & -                          & -           \\
222     & -        & FR I  & 09 36 32.019 & +04 22 10.30 & 1.339  & -   & -                          & -           \\
\hline 
\noalign{\smallskip}
223     & -        & FR II & 09 39 52.755 & +35 53 58.86 & 0.1368 & 1   & Massaro et al. (2012)      & cl          \\
223.1   & -        & FR II & 09 41 24.019 & +39 44 41.62 & 0.1075 & 1   & Massaro et al. (2010)      & -           \\
225A    & -        & UNC   & 09 42 08.480 & +13 51 54.23 & 1.565  & -   & -                          & -           \\
225B    & -        & FR II & 09 42 15.387 & +13 45 50.52 & 0.58   & -   & -                          & -           \\ 
226     & -        & FR II & 09 44 16.376 & +09 46 19.25 & 0.8177 & 1   & -                          & -           \\ 
227     & -        & FR II & 09 47 45.163 & +07 25 20.34 & 0.0858 & 2   & Hardcastle et al. (2007a)   & h           \\
228     & -        & FR II & 09 50 10.791 & +14 20 00.63 & 0.5524 & 2   & Belsole et al. (2006)      & -           \\
230     & -        & UNC   & 09 51 58.826 & -00 01 27.23 & 1.487  & -   & -                          & -           \\
231     &M\,82     & FR I  & 09 55 52.725 & +69 40 45.78 & 0.00067& 26  & -                          & -           \\
234     & -        & FR II & 10 01 49.526 & +28 47 08.87 & 0.1848 & 1   & Massaro et al. (2012)      & h           \\
\hline 
\noalign{\smallskip}
236     & -        & FR II & 10 06 01.735 & +34 54 10.43 & 0.1005 & 2   & -                          & -           \\
237     & -        & FR II & 10 08 00.030 & +07 30 16.35 & 0.877  & -   & -                          & -           \\
238     & -        & UNC   & 10 11 00.379 & +06 24 39.72 & 1.405  & -   & -                          & -           \\
239     & -        & FR II & 10 11 45.415 & +46 28 19.75 & 1.781  & -   & -                          & -           \\
241     & -        & FR II & 10 21 54.527 & +21 59 30.71 & 1.671  & 1   & Wilkes et al. (2013)       & -           \\
244.1   & -        & FR II & 10 33 33.98  & +58 14 35.70 & 0.428  & 1   & this work                  & cl          \\
245     & -        & QSR   & 10 42 44.605 & +12 03 31.26 & 1.0286 & 1   & Sambruna et al. (2004)     & -           \\
247     & -        & FR II & 10 58 58.785 & +43 01 23.12 & 0.7489 & -   & -                          & -           \\
249     & -        & QSR   & 11 02 03.848 & -01 16 17.39 & 1.554  & -   & -                          & -           \\
249.1   & -        & QSR   & 11 04 13.686 & +76 58 58.02 & 0.3115 & 1   & Stockton et al. (2006)     & -           \\
\hline 
\noalign{\smallskip}
250     & -        & FR II & 11 08 52.129 & +25 00 54.61 & ?      & -   & -                          & -           \\
252     & -        & FR II & 11 11 32.990 & +35 40 41.64 & 1.1    & 1   & Wilkes et al. (2013)       & -           \\
254     & -        & QSR   & 11 14 38.736 & +40 37 20.47 & 0.7361 & 1   & Donahue et al. (2003)      & h           \\
255     & -        & QSR   & 11 19 25.238 & -03 02 51.50 & 1.355  & -   & -                          & -           \\
256     & -        & UNC   & 11 20 43.024 & +23 27 55.22 & 1.819  & 1   & -                          & -           \\
257     & -        & UNC   & 11 23 09.172 & +05 30 19.47 & 2.474  & -   & -                          & -           \\
258     & -        & UNC   & 11 24 43.881 & +19 19 29.50 & 0.165? & 1   & Massaro et al. (2012)      & cl          \\
263     & -        & QSR   & 11 39 57.043 & +65 47 49.38 & 0.646  & 1   & Hardcastle et al. (2002b)  & h           \\
263.1   & -        & FR II & 11 43 25.084 & +22 06 56.11 & 0.824  & 1   & -                          & -           \\
264     &NGC\,3862 & FR I  & 11 45 05.009 & +19 36 22.74 & 0.0217 & 1   & Evans et al. (2006)        & cl,k        \\
\hline 
\noalign{\smallskip}
265     & -        & FR II & 11 45 28.991 & +31 33 49.43 & 0.811  & 1   & Bondi et al. (2004)        & h           \\
266     & -        & FR II & 11 45 43.367 & +49 46 08.24 & 1.275  & 1   & Wilkes et al. (2013)       & -           \\ 
267     & -        & FR II & 11 49 56.566 & +12 47 19.07 & 1.14   & 1   & Wilkes et al. (2013)       & -           \\ 
268.1   & -        & FR II & 12 00 19.210 & +73 00 45.70 & 0.97   & 1   & -                          & -           \\
268.2   & -        & FR II & 12 00 58.61  & +31 33 19.90 & 0.362  & 1   & this work                  & h           \\ 
268.3   & -        & FR II & 12 06 24.699 & +64 13 36.80 & 0.3717 & 2   & -                          & cl          \\
268.4   & -        & QSR   & 12 09 13.617 & +43 39 20.96 & 1.3978 & 1   & Wilkes et al. (2013)       & -           \\
270     &NGC\,4261 & FR I  & 12 19 23.220 & +05 49 30.77 & 0.00746& 2   & Chiaberge et al. (2003)    & cl,k        \\ 
270.1   & -        & QSR   & 12 20 33.875 & +33 43 12.05 & 1.5324 & 2   & Wilkes et al. (2013)       & -           \\
272     & -        & FR II & 12 24 28.52  & +42 06 36.3  & 0.944  & -   & -                          & -           \\
\noalign{\smallskip}
272.1   &M\,84     & FR I  & 12 25 03.743 & +12 53 13.14 & 0.0035 & 4   & Harris et al. (2002)       & cl,k        \\
273     & -        & QSR   & 12 29 06.699 & +02 03 08.59 & 0.1583 & 27  & Marshall et al. (2001)     & k           \\
274     &M\,87     & FR I  & 12h30m49.423 & +12 23 28.04 & 0.00436& 110 & Marshall et al. (2002)     & xcl,k       \\ 
274.1   & -        & FR II & 12 35 26.66  & +21 20 34.8  & 0.422  & 1   & this work                  & -           \\
275     & -        & FR II & 12 42 19.89  & -04 46 20.1  & 0.480  & 1   & this work                  & cl          \\
275.1   & -        & QSR   & 12 43 57.650 & +16 22 53.54 & 0.5551 & 1   & Crawford \& Fabian (2003)  & k,h         \\
277     & -        & UNC   & 12 51 43.580 & +50 34 24.90 & 0.414  & -   & -                          & -           \\
277.1   &Q1250+568 & QSR   & 12 52 26.349 & +56 34 19.66 & 0.3198 & 1   & Siemiginowska et al. (2008)& -           \\ 
277.2   & -        & FR II & 12 53 33.038 & +15 42 29.21 & 0.766  & -   & -                          & -           \\ 
277.3   &Coma\,A   & FR II & 12 54 12.013 & +27 37 33.94 & 0.0853 & 3   & -                          & -           \\
\noalign{\smallskip}
\hline
\end{tabular}
\end{table*} 

\begin{table*}
\tiny
\label{tab:3CR}
\begin{tabular}{rrrlllcll}
\hline
3CR  & Other & Class & R.A.    & Dec.    & z & \chn\  & Reference & X-ray \\
name &  name &       & (J2000) & (J2000) &  & obs. &           & notes \\
\hline 
\noalign{\smallskip}
280     & -        & FR II & 12 56 57.002 & +47 20 19.87 & 0.996  & 1   & Donahue et al. (2003)      & cl,h        \\  
280.1   & -        & QSR   & 13 00 33.304 & +40 09 07.72 & 1.6713 & -   & -                          & -           \\
284     & -        & FR II & 13 11 04.666 & +27 28 07.15 & 0.2394 & 1   & Massaro et al. (2012)      & -           \\
285     & -        & FR II & 13 21 17.877 & +42 35 15.16 & 0.0794 & 1   & Hardcastle et al. (2007b)  & -           \\
286     & -        & QSR   & 13 31 08.287 & +30 30 32.95 & 0.8493 & 1   & -                          & -           \\
287     &Q1328+254 & QSR   & 13 30 37.690 & +25 09 10.87 & 1.055  & 1   & Siemiginowska et al. (2008)& -           \\  
287.1   & -        & FR II & 13 32 53.257 & +02 00 45.60 & 0.2156 & 1   & Massaro et al. (2010)      & h           \\
288     & -        & FR I  & 13 38 49.87  & +38 51 09.2  & 0.246  & 1   & Lal et al. (2010)          & xcl         \\
288.1   & -        & QSR   & 13 42 13.273 & +60 21 42.85 & 0.9645 & -   & -                          & -           \\
289     & -        & FR II & 13 45 26.367 & +49 46 32.59 & 0.9674 & 1   & -                          & -           \\
\hline 
\noalign{\smallskip}
292     & -        & FR II & 13 50 42.04  & +64 29 30.6  & 0.71   & -   & -                          & -           \\ 
293     & -        & FR I  & 13 52 17.789 & +31 26 46.44 & 0.045  & 2   & Massaro et al. (2010)      & -           \\
293.1   & -        & FR II & 13 54 40.97  & +16 14 50.1  & 0.709  & -   & -                          & -           \\
294     & -        & FR II & 14 06 44.022 & +34 11 25.10 & 1.779  & 3   & Fabian et al. (2001)       & xcl,h       \\ 
295     & -        & FR II & 14 11 20.519 & +52 12 09.97 & 0.4641 & 2   & Harris et al. (2000)       & cl,h        \\
296     &NGC\,5532 & FR I  & 14 16 52.950 & +10 48 26.60 & 0.0247 & 1   & Hardcastle et al. (2005a)  & k           \\
297     & -        & QSR   & 14 17 23.999 & -04 00 47.54 & 1.4061 & -   & -                          & -           \\ 
298     &Q1416+067 & QSR   & 14 19 08.180 & +06 28 34.80 & 1.4373 & 2   & Siemiginowska et al. (2008)& -           \\  
299     & -        & FR II & 14 21 05.58  & +41 44 48.5  & 0.367  & 2   & -                          & cl          \\
300     & -        & FR II & 14 22 59.861 & +19 35 36.72 & 0.27   & 1   & Massaro et al. (2010)      & -           \\
\hline 
\noalign{\smallskip}
300.1   & -        & UNC   & 14 28 31.314 & -01 24 07.97 & 1.1588 & -   & -                          & -           \\
303     & -        & FR I  & 14 43 02.780 & +52 01 37.27 & 0.1411 & 1   & Kataoka et al. (2003b)     & cl,k,h      \\
303.1   & -        & UNC   & 14 43 14.800 & +77 07 29.00 & 0.267  & 1   & Massaro et al. (2010)      & -           \\
305     & -        & FR II & 14 49 21.661 & +63 16 14.12 & 0.0416 & 2   & Massaro et al. (2009b)     & -           \\
305.1   & -        & FR II & 14 47 09.56  & +76 56 21.8  & 1.132  & -   & -                          & -           \\
306.1   & -        & FR II & 14 55 01.40  & -04 20 59.8  & 0.441  & 1   & this work                  & cl          \\ 
309.1   &Q1458+718 & QSR   & 14 59 07.583 & +71 40 19.86 & 0.905  & 1   & Siemiginowska et al. (2008)& -           \\ 
310     & -        & FR I  & 15 04 57.10  & +26 00 56.88 & 0.0538 & 1   & Kraft et al. (2012)        & xcl         \\
313     & -        & FR II & 15 11 00.03  & +07 51 50.1  & 0.461  & 1   & this work                  & h           \\  
314.1   & -        & FR I  & 15 10 27.064 & +70 46 07.37 & 0.1197 & 1   & Massaro et al. (2012)      & cl          \\
\hline 
\noalign{\smallskip}
315     & -        & FR I  & 15 13 40.054 & +26 07 30.06 & 0.1083 & 1   & Massaro et al. (2010)      & cl          \\
317     & -        & FR I  & 15 16 44.489 & +07 01 17.85 & 0.0344 & 11  & Blanton et al. (2001)      & xcl         \\
318     & -        & FR II & 15 20 05.448 & +20 16 05.76 & 1.574  & 1   & Wilkes et al. (2013)       & -           \\ 
318.1   &NGC\,5920 & UNC   & 15 21 51.851 & +07 42 31.75 & 0.0453 & 1   & Mazzotta et al. (2002)     & xcl         \\
319     & -        & FR II & 15 24 05.640 & +54 28 18.40 & 0.192  & 1   & Massaro et al. (2012)      & cl          \\
320     & -        & FR II & 15 31 25.37  & +35 33 40.0  & 0.342  & 1   & this work                  & cl          \\
321     & -        & FR II & 15 31 43.512 & +24 04 18.82 & 0.0961 & 2   & Hardcastle et al. (2004)   & h           \\ 
322     & -        & FR II & 15 35 01.230 & +55 36 52.87 & 1.681  & -   & -                          & -           \\ 
323     & -        & UNC   & 15 41 45.534 & +60 15m35.07 & 0.679  & -   & -                          & -           \\
323.1   & -        & QSR   & 15 47 43.545 & +20 52 16.54 & 0.2643 & 1   & Massaro et al. (2010)      & cl          \\
\hline 
\noalign{\smallskip}
324     & -        & FR II & 15 49 48.897 & +21 25 38.06 & 1.2063 & 1   & -                          & cl           \\
325     & -        & FR II & 15 49 58.424 & +62 41 21.66 & 1.135  & 2   & Hardcastle et al. (2009)   & -           \\                           
326     & -        & FR II & 15 52 09.14  & +20 05 35.8  & 0.0895 & 3   & -                          & -           \\
326.1   & -        & FR II & 15 56 10.068 & +20 04 20.44 & 1.825  & -   & -                          & -           \\ 
327     & -        & FR II & 16 02 27.375 & +01 57 56.16 & 0.1048 & 1   & Hardcastle et al. (2007a)  & cl          \\
327.1   & -        & UNC   & 16 04 45.38  & +01 17 50.3  & 0.462  & 1   & this work                  & k           \\ 
330     & -        & FR II & 16 09 36.607 & +65 56 43.61 & 0.55   & 1   & Hardcastle et al. (2002b)  & cl          \\
332     & -        & FR II & 16 17 42.540 & +32 22 34.49 & 0.1515 & 1   & Massaro et al. (2010)      & cl          \\
334     & -        & QSR   & 16 20 21.819 & +17 36 24.01 & 0.5551 & 1   & Hardcastle et al. (2002a)  & -           \\
336     & -        & QSR   & 16 24 39.086 & +23 45 12.20 & 0.9273 & 1   & -                          & -           \\
\hline 
\noalign{\smallskip}
337     & -        & FR II & 16 28 52.846 & +44 19 05.08 & 0.635  & 1   & -                          & cl          \\
338     &NGC\,6166 & FR I  & 16 28 38.480 & +39 33 05.60 & 0.03035& 6   & Johnstone et al. (2002)    & xcl         \\ 
340     & -        & FR II & 16 29 36.932 & +23 20 14.42 & 0.7754 & 1   & -                          & -           \\
341     & -        & FR II & 16 28 03.98  & +27 41 39.3  & 0.448  & 1   & this work                  & k           \\
343     & -        & QSR   & 16 34 33.789 & +62 45 35.81 & 0.988  & 1   & -                          & -           \\
343.1   & -        & FR II & 16 38 28.193 & +62 34 44.31 & 0.75   & 1   & -                          & -           \\ 
345     & 1641+399 & QSR   & 16 42 58.809 & +39 48 36.99 & 0.5928 & 3   & Sambruna et al. (2004)     & k           \\
346     & -        & FR I  & 16 43 48.599 & +17 15 49.46 & 0.162  & 1   & Donato et al. (2004)       & k           \\
348     &Hercules A& FR I  & 16 51 08.147 & +04 59 33.32 & 0.154  & 3   & Nulsen et al. (2005)       & xcl         \\
349     & -        & FR II & 16 59 28.893 & +47 02 55.04 & 0.205  & 1   & Massaro et al. (2010)      & h           \\
\noalign{\smallskip}
\hline
\end{tabular}
\end{table*} 

\begin{table*}
\tiny
\label{tab:3CR}
\begin{tabular}{rrrlllcll}
\hline
3CR  & Other & Class & R.A.    & Dec.    & z & \chn\  & Reference & X-ray \\
name &  name &       & (J2000) & (J2000) &  & obs. &           & notes \\
\hline 
\noalign{\smallskip}
351     & -        & FR II & 17 04 41.376 & +60 44 30.50 & 0.3719 & 2   & Brunetti et al. (2001)     & h           \\
352     & -        & FR II & 17 10 44.106 & +46 01 28.56 & 0.8067 & 1   & -                          & -           \\
353     & -        & FR II & 17 20 28.158 & -00 58 46.62 & 0.0304 & 2   & Kataoka et al. (2008)      & k           \\
356     & -        & FR II & 17 24 19.041 & +50 57 40.14 & 1.079  & 1   & Wilkes et al. (2013)       & -           \\
357     & -        & FR II & 17 28 20.109 & +31 46 02.58 & 0.1662 & 1   & Massaro et al. (2012)      & cl          \\
368     & -        & FR II & 18 05 06.454 & +11 01 35.06 & 1.131  & 1   & Wilkes et al. (2013)       & -           \\
371     & -        & BL    & 18 06 50.680 & +69 49 28.10 & 0.051  & 2   & Pesce et al. (2001)        & k           \\
379.1   & -        & FR II & 18 24 32.976 & +74 20 59.00 & 0.256  & 1   & Massaro et al. (2012)      & -           \\
380     & 1828+487 & QSR   & 18 29 31.780 & +48 44 46.16 & 0.692  & 1   & Marshall et al. (2005)     & k           \\
381     & -        & FR II & 18 33 46.301 & +47 27 02.61 & 0.1605 & 1   & Massaro et al. (2010)      & -           \\
\hline 
\noalign{\smallskip}
382     & -        & FR II & 18 35 03.390 & +32 41 46.80 & 0.0578 & 2   & Gliozzi et al. (2007)      & -           \\
386     & -        & FR I  & 18 38 26.255 & +17 11 49.28 & 0.0168 & 1   & -                          & -           \\
388     & -        & FR II & 18 44 02.40  & +45 33 29.7  & 0.0917 & 2   & Hodges-Kluck et al. (2010) & cl          \\
389     & -        & UND   & 18 45 41.50  & -03 07 01.92 & ?      & -   & -                          & -           \\
390     & -        & UND   & 18 45 37.621 & +09 53 44.71 & ?      & -   & -                          & -           \\
390.3   & -        & FR II & 18 42 08.989 & +79 46 17.12 & 0.0561 & 1   & Hardcastle et al. (2007a)  & k,h         \\
394     & -        & UND   & 18 59 23.362 & +12 59 12.09 & ?      & -   & -                          & -           \\
399.1   & -        & FR II & 19 15 56.763 & +30 19 53.79 & ?      & -   & -                          & -           \\ 
401     & -        & FR II & 19 40 25.117 & +60 41 34.94 & 0.2011 & 2   & Reynolds et al. (2005)     & xcl         \\ 
402     & -        & FR I  & 19 41 45.899 & +50 35 45.86 & 0.0239 & 1   & Massaro et al. (2012)      & cl,k        \\
\hline 
\noalign{\smallskip}
403     & -        & FR II & 19 52 15.809 & +02 30 24.18 & 0.059  & 1   & Kraft et al. (2005)        & k,h         \\
403.1   & -        & FR II & 19 52 30.50  & -01 17 18.00 & 0.0554 & 1   & Massaro et al. (2012)      & cl          \\
405     & Cygnus A & FR II & 19 59 28.356 & +40 44 02.09 & 0.056  & 11  & Wilson et al. (2002)       & xcl,h       \\ 
409     & -        & FR II & 20 14 27.596 & +23 34 52.91 & ?      & -   & -                          & -           \\ 
410     & -        & FR II & 20 20 06.60  & +29 42 14.20 & 0.2485 & 1   & Massaro et al. (2012)      & -           \\
411     & -        & FR II & 20 22 08.44  & +10 01 11.30 & 0.467  & 1   & this work                  & -           \\
415.2   & -        & UND   & 20 32 46.115 & +53 45 50.14 & ?      & -   & -                          & -           \\
418     & -        & QSR   & 20 38 37.034 & +51 19 12.66 & 1.686  & -   & -                          & -           \\
424     & -        & FR I  & 20 48 12.087 & +07 01 17.17 & 0.127  & 1   & Massaro et al. (2012)      & cl          \\
427.1   & -        & FR II & 21 04 06.865 & +76 33 10.82 & 0.572  & 1   & Hardcastle et al. (2004)   & -           \\
\hline 
\noalign{\smallskip}
428     & -        & UND   & 21 08 22.39  & +49 36 37.6  & ?      & -   & -                          & -           \\
430     & -        & FR II & 21 18 19.094 & +60 48 07.77 & 0.0541 & 1   & Massaro et al. (2012)      & cl          \\
431     & -        & UND   & 21 18 52.580 & +49 36 58.80 & ?      & -   & -                          & -           \\
432     & -        & QSR   & 21 22 46.329 & +17 04 37.95 & 1.785  & 1   & Erlund et al. (2006)       & -           \\
433     & -        & FR I  & 21 23 44.534 & +25 04 11.88 & 0.1016 & 1   & Miller \& Brandt (2009)    & -           \\ 
434     & -        & FR II & 21 23 16.24  & +15 48 05.80 & 0.322  & 1   & this work                  & cl          \\
435B    & -        & FR II & 21 29 06.10  & +07 32 54.80 & 0.865  & 1   & this work                  & -           \\
436     & -        & FR II & 21 44 11.743 & +28 10 18.91 & 0.2145 & 2   & Massaro et al. (2012)      & h           \\
437     & -        & FR II & 21 47 25.106 & +15 20 37.49 & 1.48   & 1   & Wilkes et al. (2013)       & -           \\
438     & -        & FR II & 21 55 52.290 & +38 00 29.62 & 0.29   & 3   & Hardcastle et al. (2004)   & xcl         \\ 
\hline 
\noalign{\smallskip}
441     & -        & FR II & 22 06 04.916 & +29 29 19.98 & 0.708  & 1   & -                          & -           \\ 
442     & -        & FR I  & 22 14 46.90  & +13 50 24.00 & 0.0263 & 4   & Hardcastle et al. (20007b) & xcl         \\
445     & -        & FR II & 22 23 49.530 & -02 06 12.85 & 0.0558 & 8   & Perlman et al. (2010)      & cl,h        \\
449     & -        & FR I  & 22 31 20.9   & +39 21 48.00 & 0.0171 & 3   & Evans et al. (2006)        & cl          \\
452     & -        & FR II & 22 45 48.77  & +39 41 15.99 & 0.0811 & 1   & Isobe et al. (2002)        & h,l         \\
454     & -        & QSR   & 22 51 34.736 & +18 48 40.12 & 1.757  & -   & -                          & -           \\ 
454.1   & -        & FR II & 22 50 32.937 & +71 29 19.18 & 1.841  & -   & -                          & cl          \\
454.2   & -        & UND   & 22 52 05.41  & +64 40 10.42 & ?      & -   & -                          & -           \\
454.3   & -        & QSR   & 22 53 57.747 & +16 08 53.56 & 0.859  & 5   & Marshall et al. (2005)     & k           \\
455     & -        & QSR   & 22 55 03.889 & +13 13 33.99 & 0.543  & 1   & -                          & -           \\
\hline 
\noalign{\smallskip}
456     & -        & FR II & 23 12 28.076 & +09 19 26.39 & 0.2330 & 1   & Massaro et al. (2012)      & -           \\
458     & -        & FR II & 23 12 52.083 & +05 16 49.77 & 0.289  & 1   & Massaro et al. (2012)      & cl,h        \\
459     & -        & FR II & 23 16 35.30  & +04 05 18.30 & 0.2199 & 1   & Massaro et al. (2012)      & l           \\
460     & -        & FR II & 23 21 28.510 & +23 46 48.45 & 0.268  & 1   & Massaro et al. (2010)      & cl          \\
465     &NGC\,7720 & FR I  & 23 38 29.385 & +27 01 53.25 & 0.0302 & 1   & Hardcastle et al. (2005b)  & xcl,k       \\
468.1   & -        & UND   & 23 50 54.849 & +64 40 19.54 & ?      & -   & -                          & -           \\
469.1   & -        & FR II & 23 55 23.32  & +79 55 19.60 & 1.336  & 1   & Wilkes et al. (2013)       & -           \\ 
470     & -        & FR II & 23 58 35.890 & +44 04 45.55 & 1.653  & 1   & Wilkes et al. (2013)       & -           \\
\noalign{\smallskip}
\hline
\end{tabular}
\end{table*}

\section{B. Images of the sources}
\label{sec:appendixB}
Although for many radio sources analyzed the X-ray data are comprised of
rather few counts, the radio morphologies are shown here via contour
diagrams which are superposed on X-ray event files that have been
smoothed with a Gaussian.  The full width half maximum (FWHM) of the
Gaussian smoothing function is given in the figure captions.  When there is
sufficient signal to noise ratio (S/N) of the X-ray image to provide spatial information, 
contours were added (cyan or white)
which are normally separated by factors of two.  Most of the overlaid
radio contours increase by factors of four.
The X-ray event files shown are in units of counts/pixel in the 0.5-7 keV energy range.
The primary reason figures appear so different from each other is the
wide range in angular size of the radio sources.  

\begin{figure}
\includegraphics[keepaspectratio=true,scale=0.90]{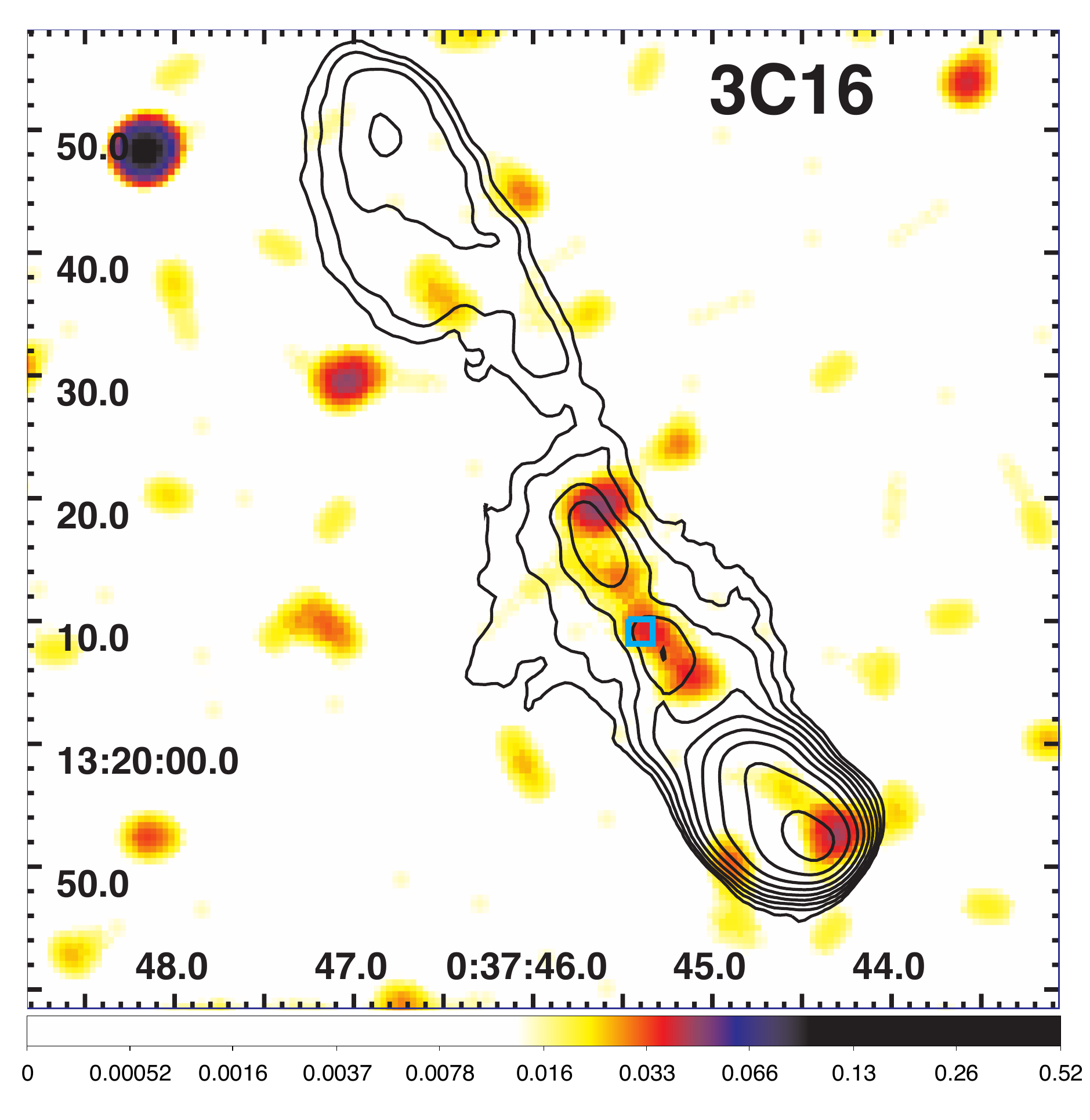}
\caption{The X-ray image of 3C\,16 for the energy band 0.5-7 keV.  The
  event file has not been regridded so the pixel size is 0.492\arcsec.
  The image has been smoothed with a Gaussian of FWHM=4\arcsec.  Since
  there is only a marginal detection of an X-ray nucleus, the map has not been
  registered.  The small square (cyan or black) marks the location of
  the optical identification: a faint galaxy with a companion
  \citep{riley80}.  The radio contours (black) come from an 8.4 GHz
  map kindly supplied by M. J. Hardcastle \citep{gilbert04} and start
  at 0.1 mJy/beam, increasing by factors of two.  The clean beam is
  2.5\arcsec.}
\label{fig:3c16app}
\end{figure}

\begin{figure}
\includegraphics[keepaspectratio=true,scale=0.90]{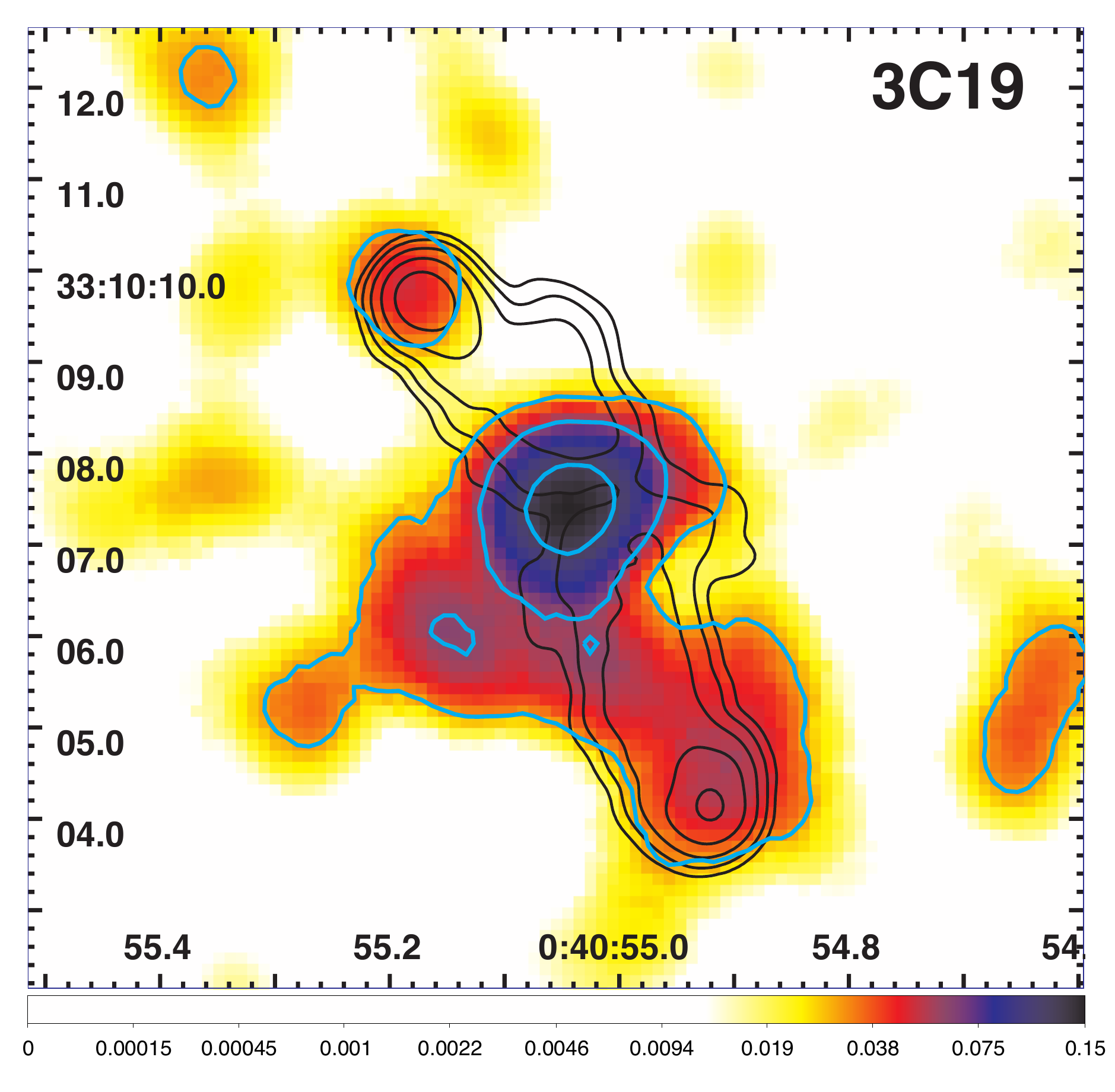}
\caption{The X-ray image of 3C\,19 for the energy band 0.5-7 keV.  The
  event file has been regridded to a pixel size of 0.123\arcsec\ and
  smoothed with a Gaussian of FWHM=1.3\arcsec.  X-ray contours (white
  or cyan) start at 0.03 counts/pix and increase by factors of two.
  The radio contours (black) come from a 4.86 GHz map kindly supplied
  by M. Hardcastle \citep{gilbert04} and start at 0.4 mJy/beam,
  increasing by factors of four.  The clean beam is 0.4\arcsec.}
\label{fig:3c19app}
\end{figure}

\begin{figure}
\includegraphics[keepaspectratio=true,scale=0.90]{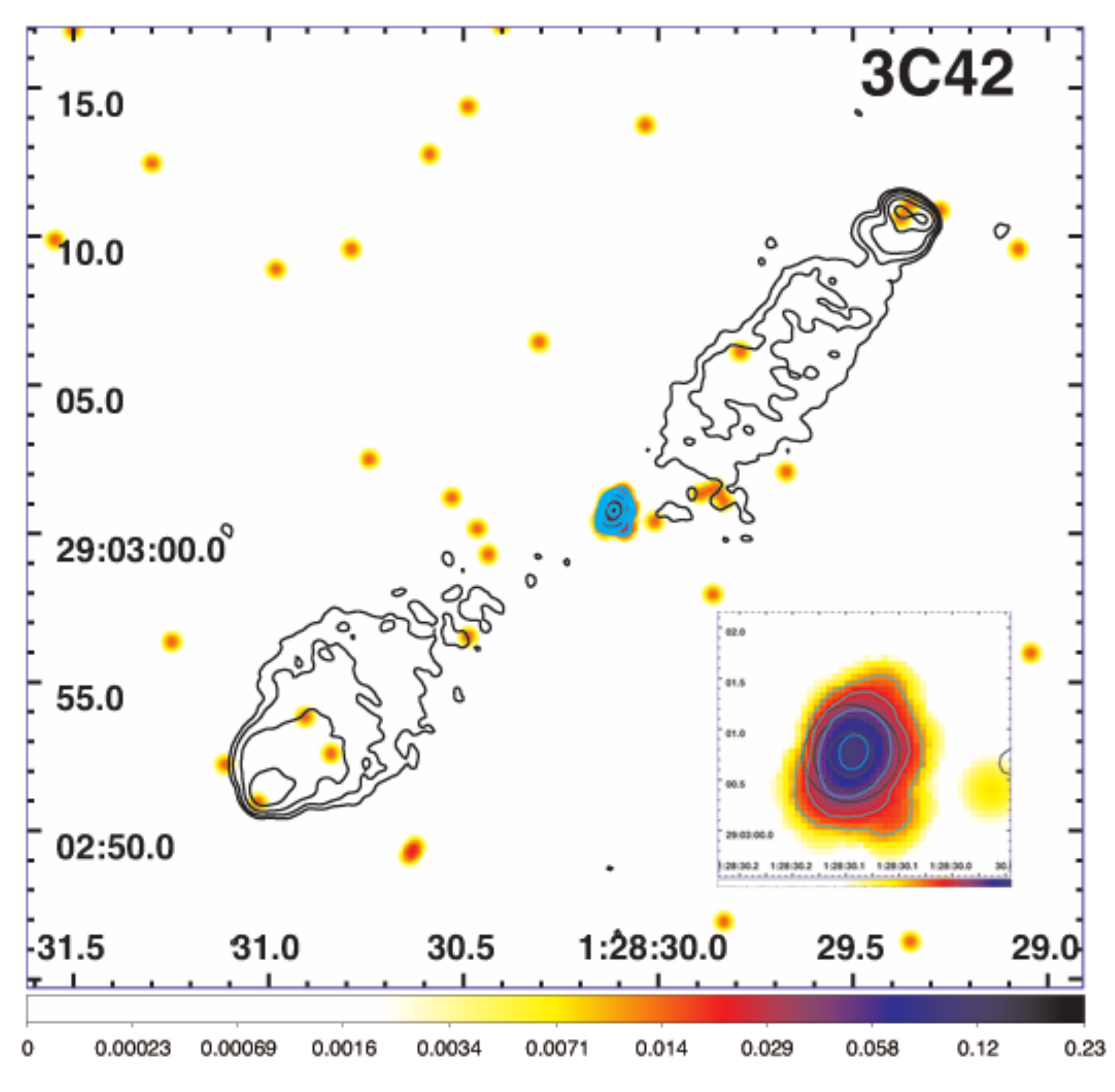}
\caption{The X-ray image of 3C\,42 for the energy band 0.5-7 keV.  The
  event file has been regridded to a pixel size of 0.0615\arcsec\ and
  smoothed with a Gaussian of FWHM=0.5\arcsec.  X-ray contours (white
  or cyan) start at 0.025 counts/pix and increase by factors of two.
  The radio contours (black) come from an 8.4 GHz map kindly supplied
  by M. Hardcastle \citep{gilbert04} and start at 0.1 mJy/beam,
  increasing by factors of four. The clean beam is 0.4\arcsec\ FWHM.
  The insert shows an enlarged version of the nucleus.}
\label{fig:3c42app}
\end{figure}

\begin{figure}
\includegraphics[keepaspectratio=true,scale=0.90]{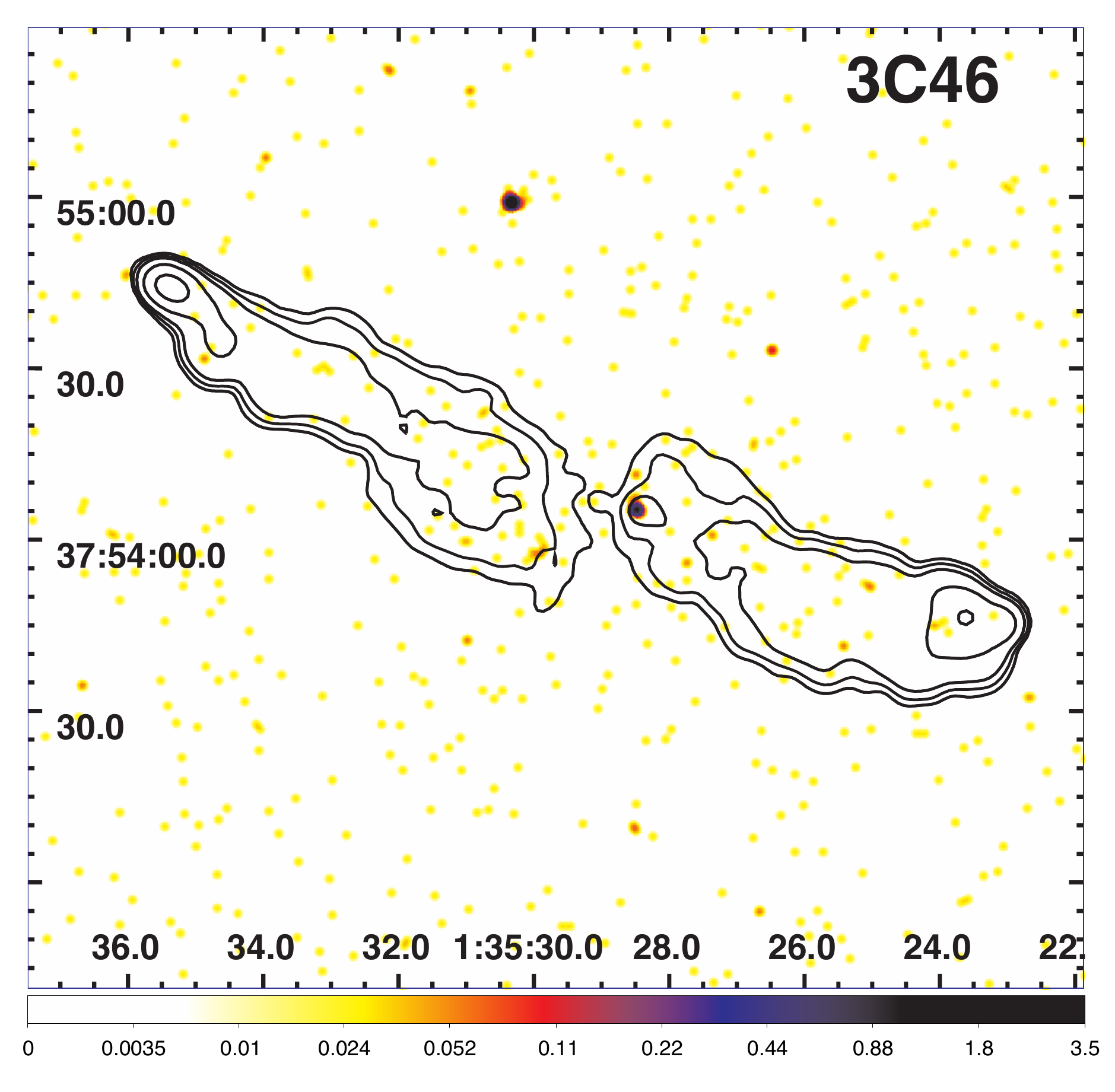}
\caption{The X-ray image of 3C\,46 for the energy band 0.5-7 keV.  The event file
has been regridded to a pixel size of 0.246\arcsec\ and
smoothed with a Gaussian of FWHM=1.45\arcsec. The radio
contours (black) come from a 1.5 GHz map from the DRAGN website and
downloaded from NED.  The contours start at 1 mJy/beam, increasing by
factors of four (but with an extra contour at 2 mJy/beam).  The clean
beam is 4.2\arcsec.}
\label{fig:3c46app}
\end{figure}

\begin{figure}
\includegraphics[keepaspectratio=true,scale=0.90]{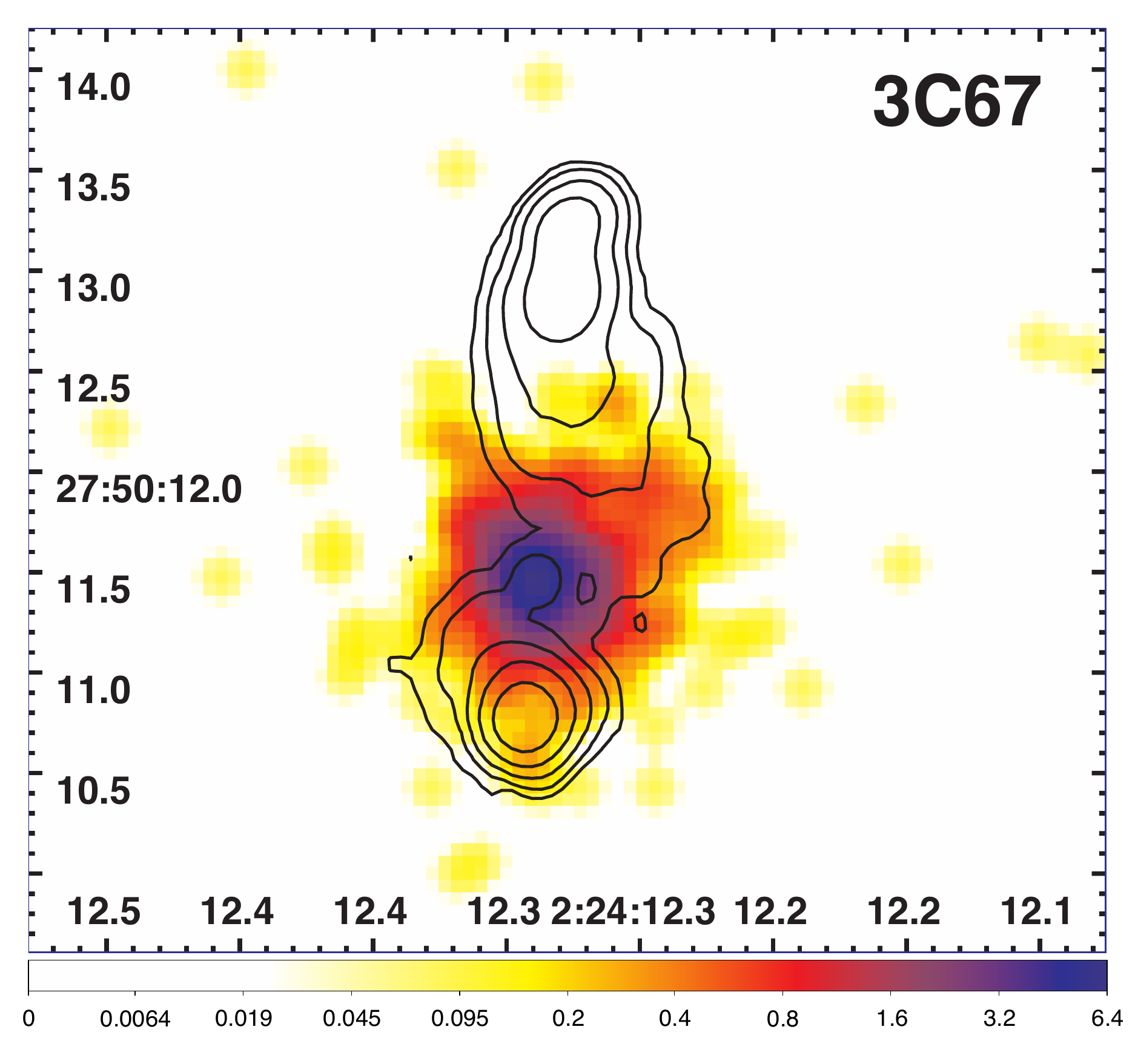}
\caption{The X-ray image of 3C\,67 for the energy band 0.5-7 keV.  The
  event file has been regridded to a pixel size of 0.0615\arcsec\ and
  smoothed with a Gaussian of FWHM=0.22\arcsec.  The radio contours
  (black) come from an 8.4 GHz VLA map kindly supplied by
  M. Hardcastle \citep{gilbert04} and start at 0.25 mJy/beam,
  increasing by factors of four.  The clean beam is 0.2\arcsec.}
\label{fig:3c67app}
\end{figure}

\begin{figure}
\includegraphics[keepaspectratio=true,scale=0.80]{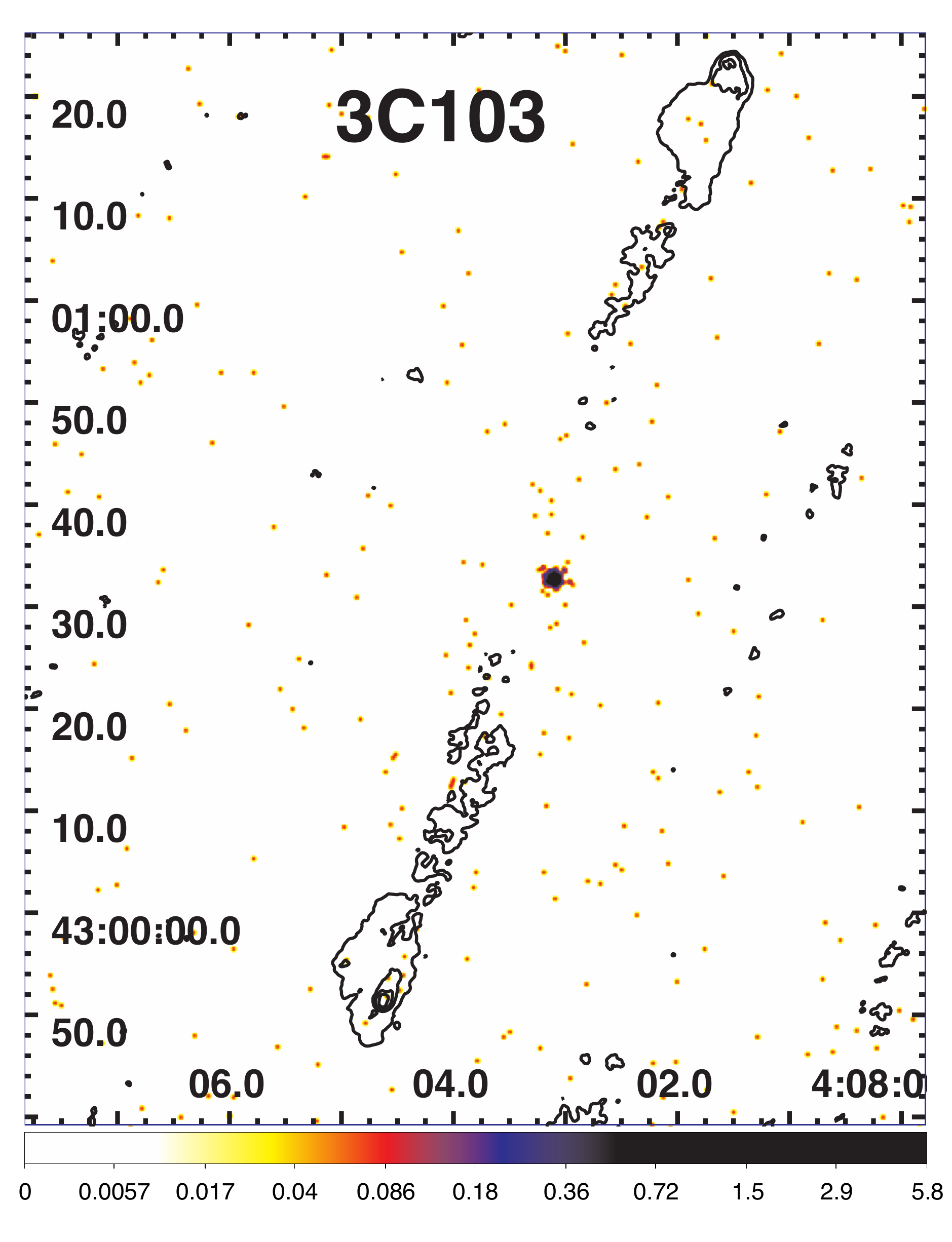}
\caption{The X-ray image of 3C\,103 for the energy band 0.5-7 keV.  The event
file has been regridded to a pixel size of 0.123\arcsec\ and
smoothed with a Gaussian of FWHM=0.43\arcsec.  The radio
contours (black) come from an 8.44 GHz VLA map kindly supplied by
C. C. Cheung, and start at 0.3 mJy/beam, increasing by factors of
four.  The clean beam is 0.37\arcsec.}
\label{fig:3c103app}
\end{figure}

\begin{figure}
\includegraphics[keepaspectratio=true,scale=0.90]{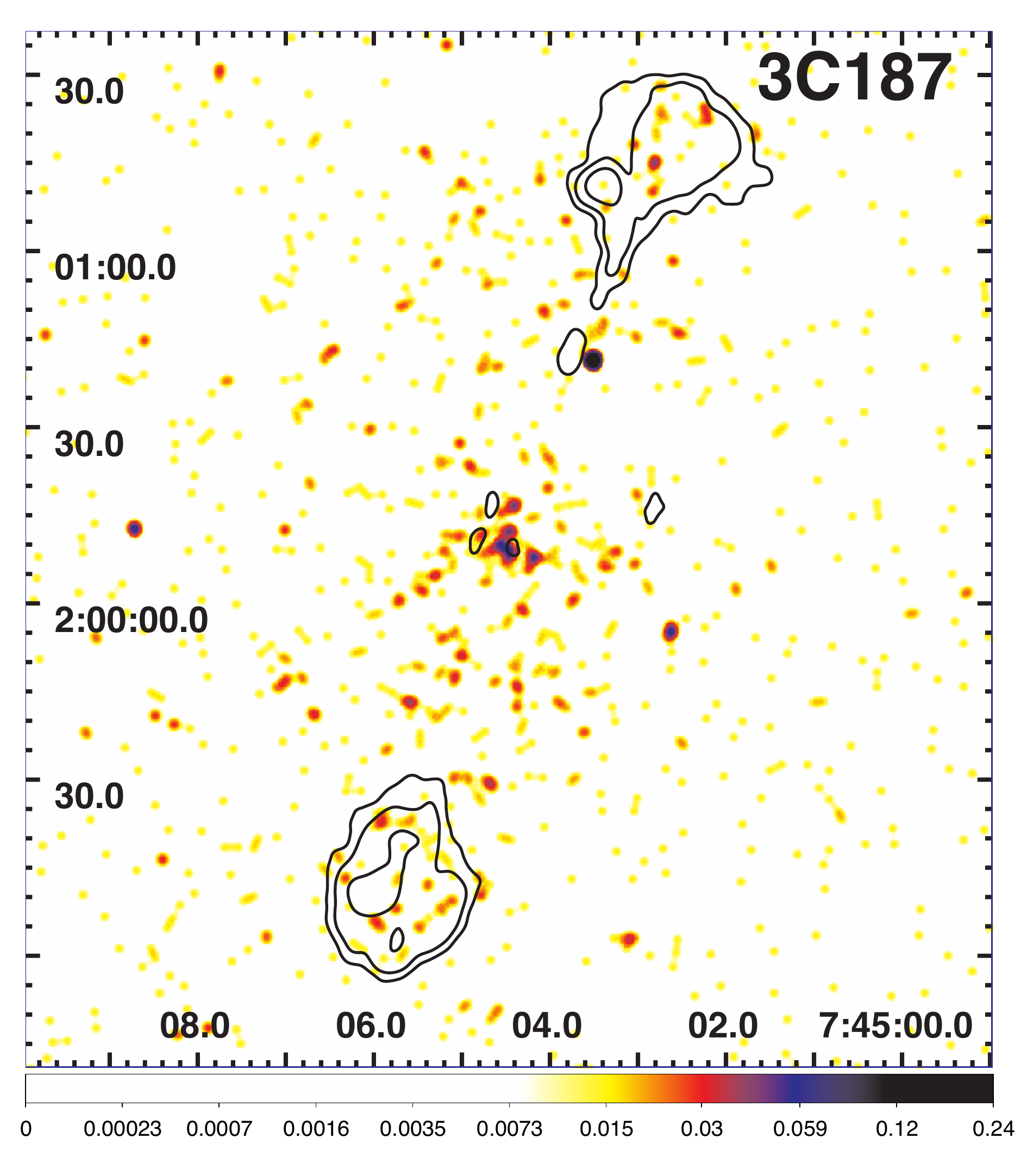}
\caption{The X-ray image of 3C\,187 for the energy band 0.5-7 keV.  The event
file has been regridded to a pixel size of 0.246\arcsec\ and
smoothed with a Gaussian of FWHM=2.0\arcsec.  The radio
contours (black) come from a 1.4 GHz map constructed from archival u,v
data, and start at 1 mJy/beam, increasing by factors of four.  The
clean beam is 3.0\arcsec.  Registration of the X-ray image
is approximate since there is no well defined X-ray nucleus.
The VLA image at 1.4 GHz have been obtained via AIPS standard reduction procedure (http://www.aips.nrao.edu/cook.html).}
\label{fig:3c187app}
\end{figure}

\begin{figure}
\includegraphics[keepaspectratio=true,scale=0.90]{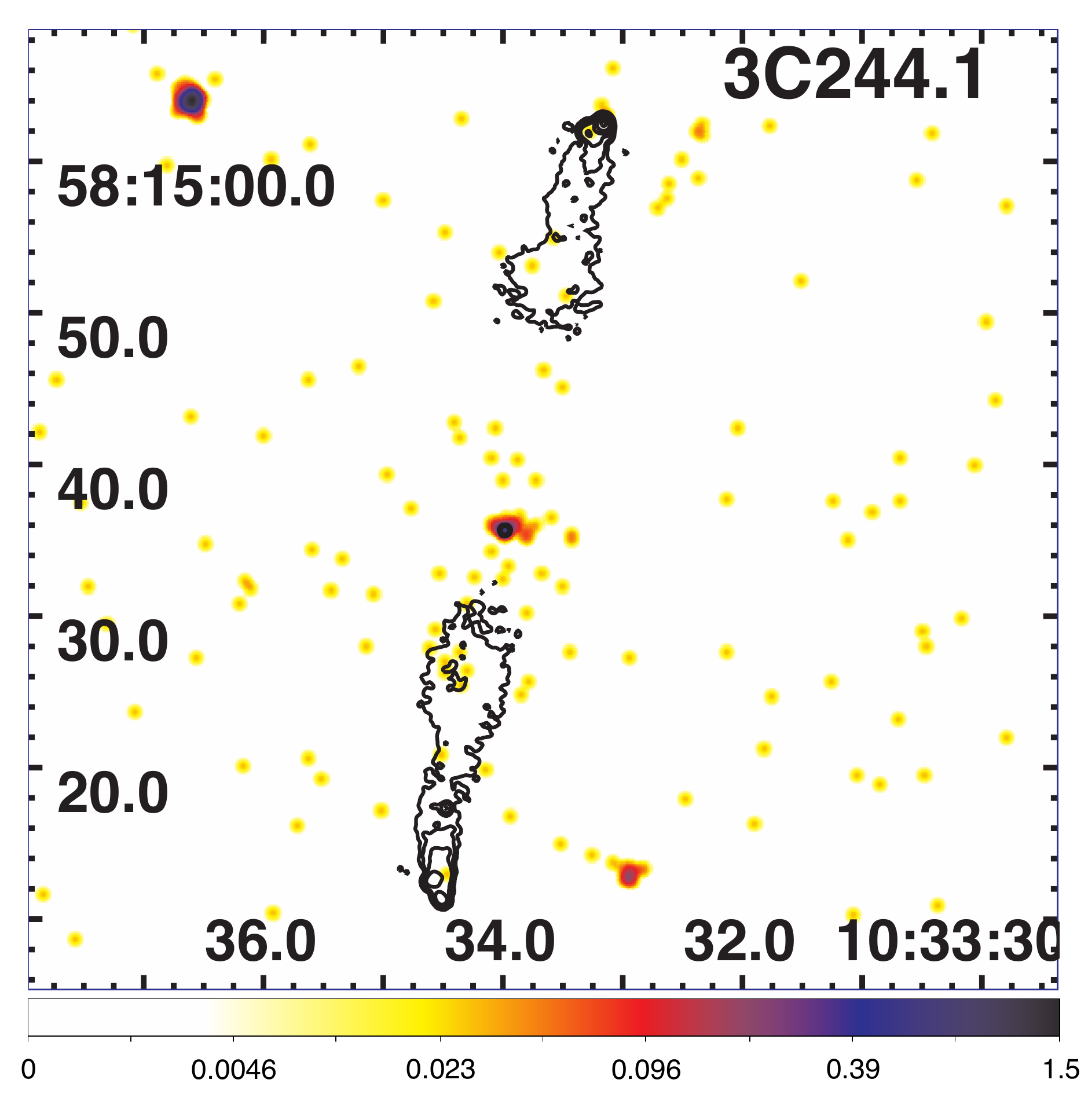}
\caption{The X-ray image of 3C\,244.1 for the energy band 0.5-7 keV.
  The event file has been regridded to a pixel size of
  0.123$^{\prime\prime}$ and smoothed with a Gaussian of
  FWHM=0.72$^{\prime\prime}$.  The radio contours (black) come from a
  8.45 GHz map kindly supplied by M. Hardcastle \citep{gilbert04} and
  start at 0.1 mJy/beam, increasing by factors of four.  The clean
  beam is 0.4$^{\prime\prime}$.}
\label{fig:3c244.1app}
\end{figure}

\begin{figure}
\includegraphics[keepaspectratio=true,scale=0.90]{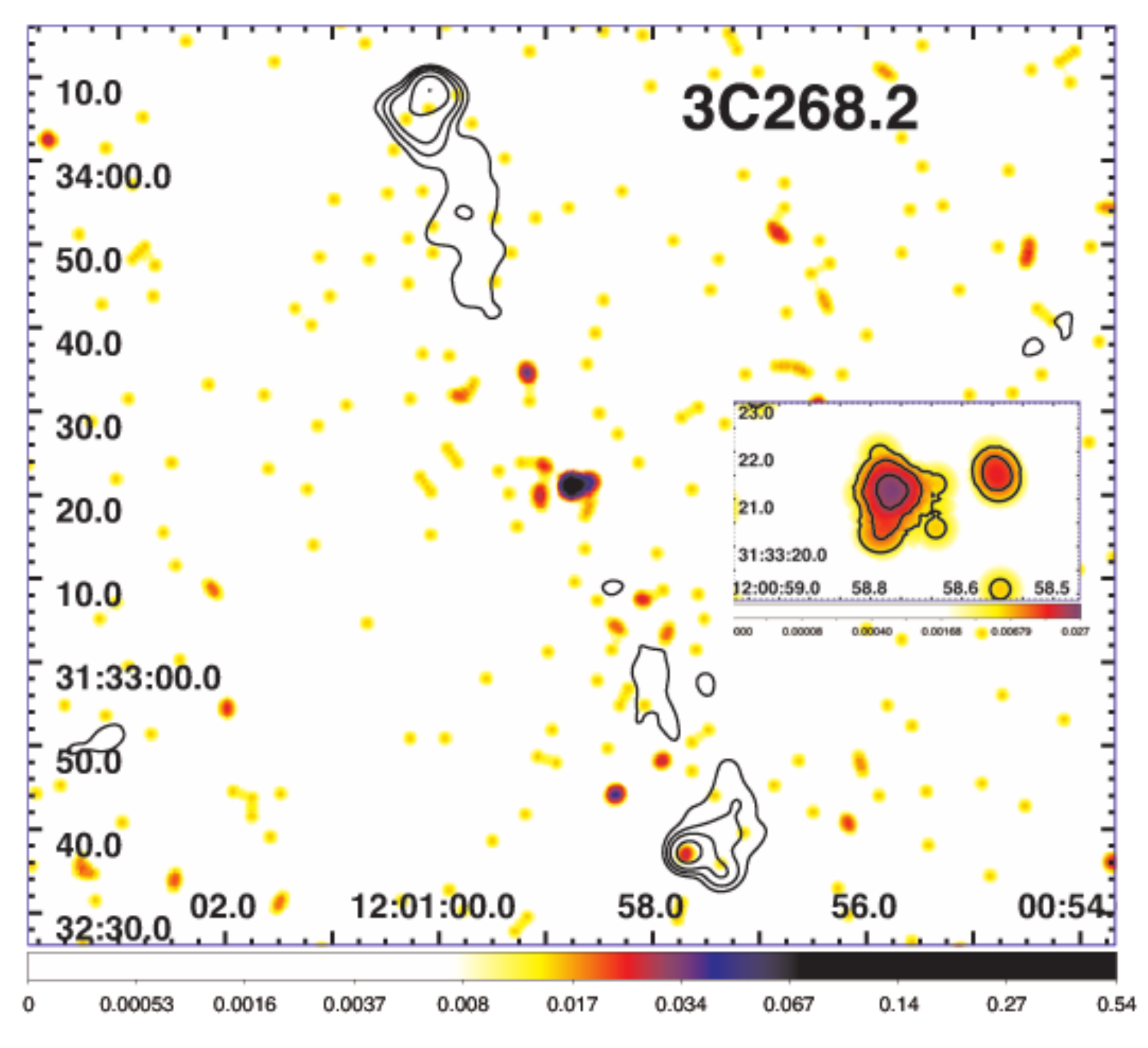}
\caption{The X-ray image of 3C\,268.2 for the energy band 0.5-7 keV.
  The event file has been regridded to a pixel size of
  0.246\arcsec\ and smoothed with a Gaussian of FWHM=2.0\arcsec.  The
  radio contours (black) come from a 1.4 GHz map downloaded from NED
  and start at 5 mJy/beam, increasing by factors of four.  The clean
  beam is 2.7\arcsec\ x 2.5\arcsec\ with major axis in
  PA=-69$^{\circ}$.  Since no radio nucleus was detected, the X-ray
  image has not been registered.  The insert shows an enlarged version
  of the nuclear region.  In this case, the event file has been
  regridded to a pixel size of 0.0615$^{\prime\prime}$ and smoothed
  with a Gaussian of FWHM=0.8$^{\prime\prime}$.  X-ray contours (white
  or black) are 0.005, 0.01, and 0.02 counts/pix. }
\label{fig:3c268.2app}
\end{figure}

\begin{figure}
\includegraphics[keepaspectratio=true,scale=0.90]{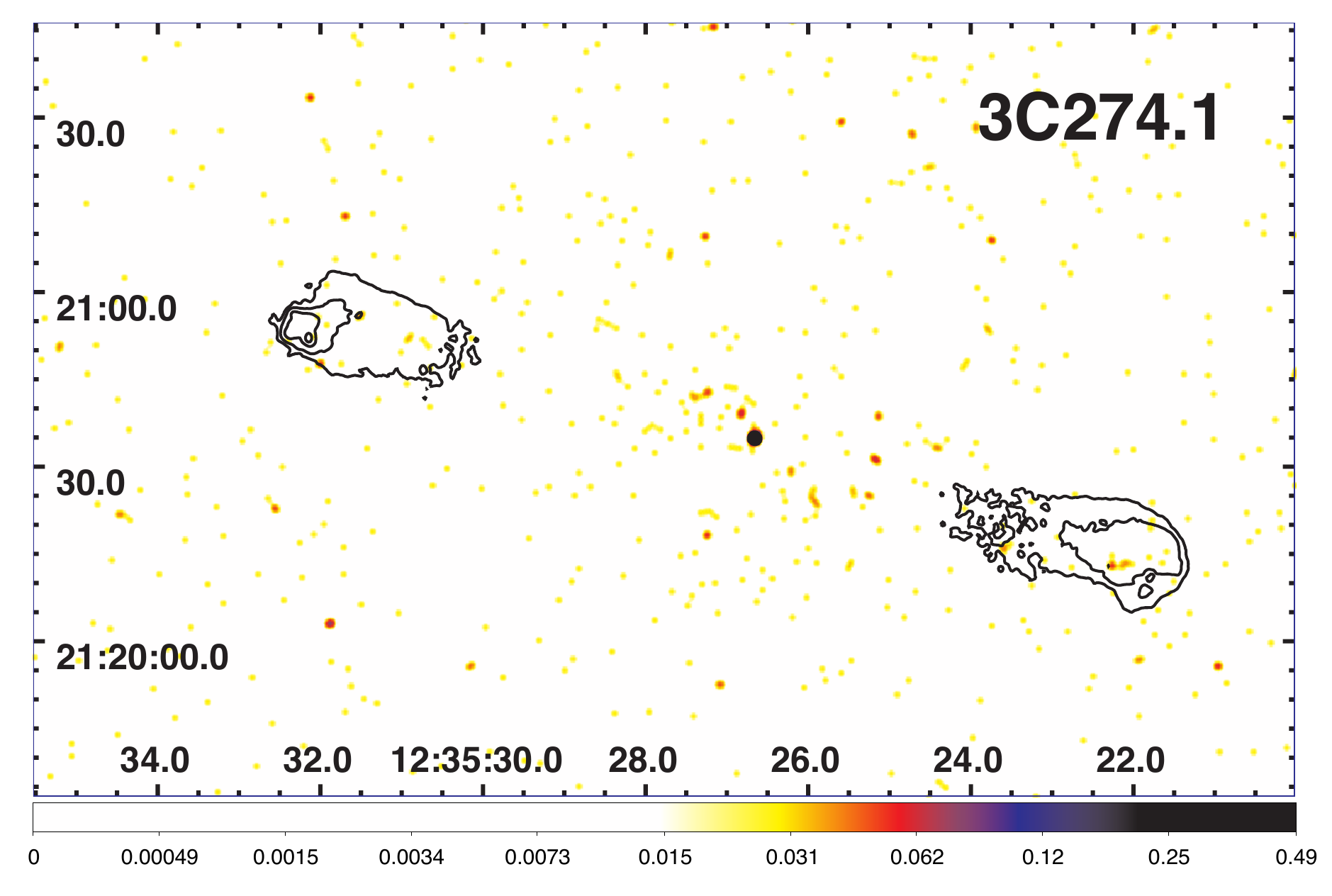}
\caption{The X-ray image of 3C\,274.1 for the energy band 0.5-7 keV.
  The event file has been regridded to a pixel size of
  0.246\arcsec\ and smoothed with a Gaussian of FWHM=1.45\arcsec.  The
  radio contours (black) come from an 8.5 GHz map kindly supplied by
  M. Hardcastle \citep{gilbert04} and start at 0.125 mJy/beam,
  increasing by factors of four.  The clean beam is 1.0\arcsec.}
\label{fig:3c274.1app}
\end{figure}

\begin{figure}
\includegraphics[keepaspectratio=true,scale=0.90]{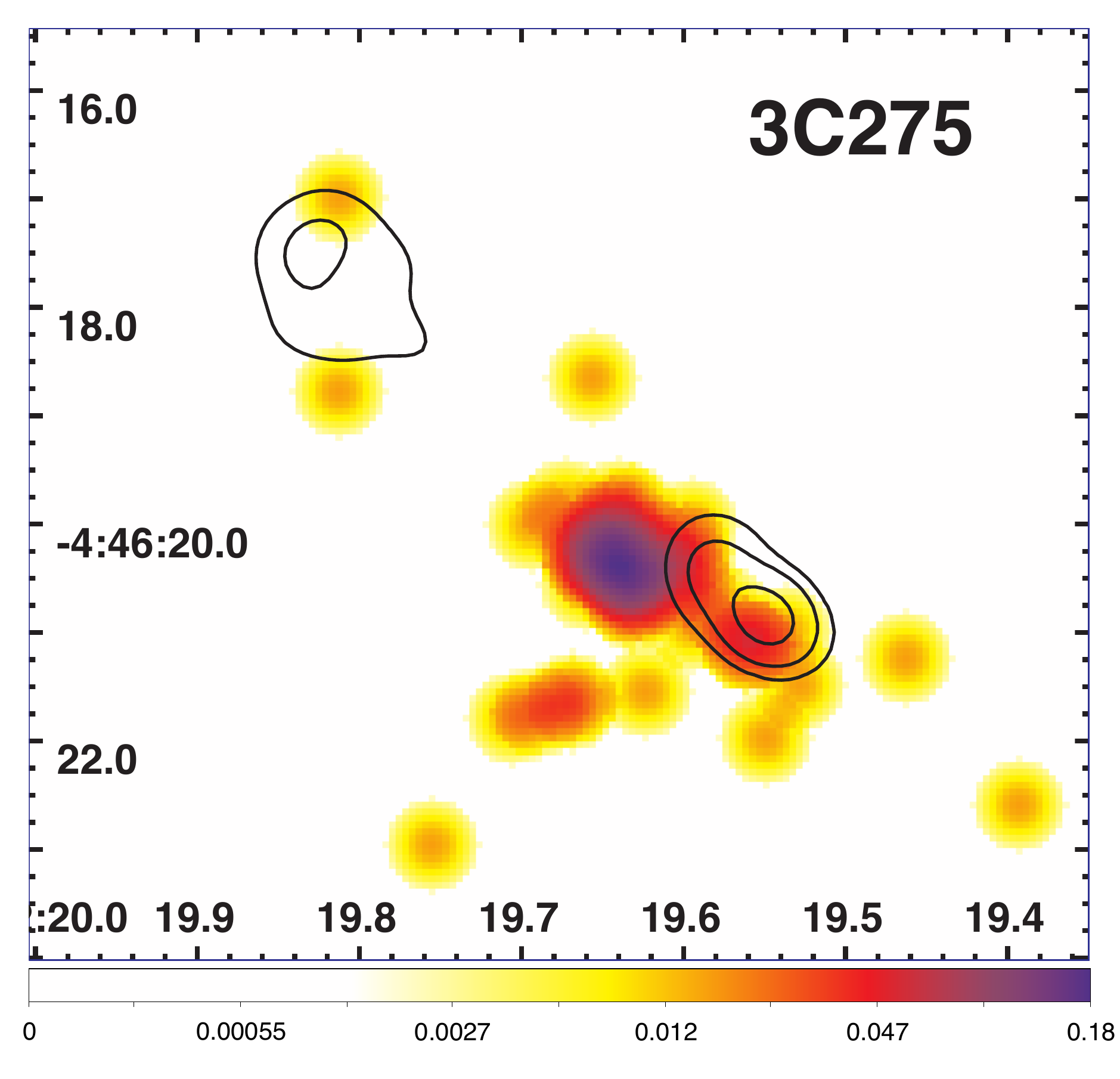}
\caption{The X-ray image of 3C\,275 for the energy band 0.5-7 keV.  The event
file has been regridded to a pixel size of 0.0615\arcsec\ and
smoothed with a Gaussian of FWHM=0.5\arcsec.  The radio
contours (black) come from an 8.4 GHz map downloaded from the NVAS and
start at 2 mJy/beam, increasing by factors of four.  The clean beam is
0.34\arcsec\ x 0.24\arcsec\ with major axis in
PA=52$^{\circ}$.  No clear detection of the radio nucleus has been
found on the maps available to us (L,C,X, and U bands); thus the X-ray
map has not been registered.}
\label{fig:3c275app}
\end{figure}

\begin{figure}
\includegraphics[keepaspectratio=true,scale=0.90]{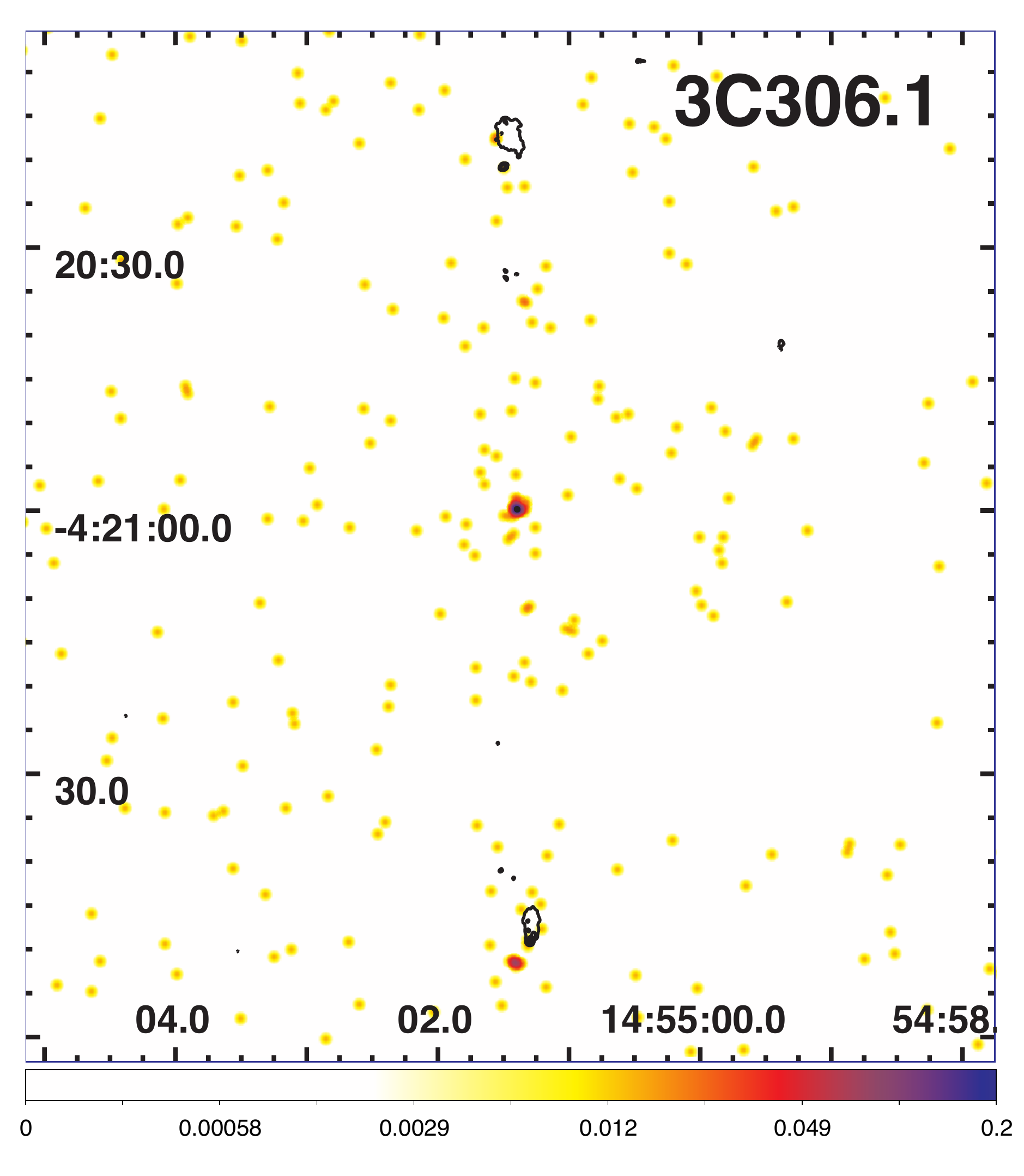}
\caption{The X-ray image of 3C\,306.1 for the energy band 0.5-7 keV.  The event
file has been regridded to a pixel size of 0.123\arcsec\ and
smoothed with a Gaussian of FWHM=1\arcsec.  The radio
contours (black) come from an 8.4 GHz VLA map kindly supplied by
C. C. Cheung, and start at 0.5 mJy/beam, increasing by factors of
four.  The clean beam is 0.31\arcsec.  There are two counts
aligned with the S hotspot.}
\label{fig:3c306.1app}
\end{figure}

\begin{figure}
\includegraphics[keepaspectratio=true,scale=0.90]{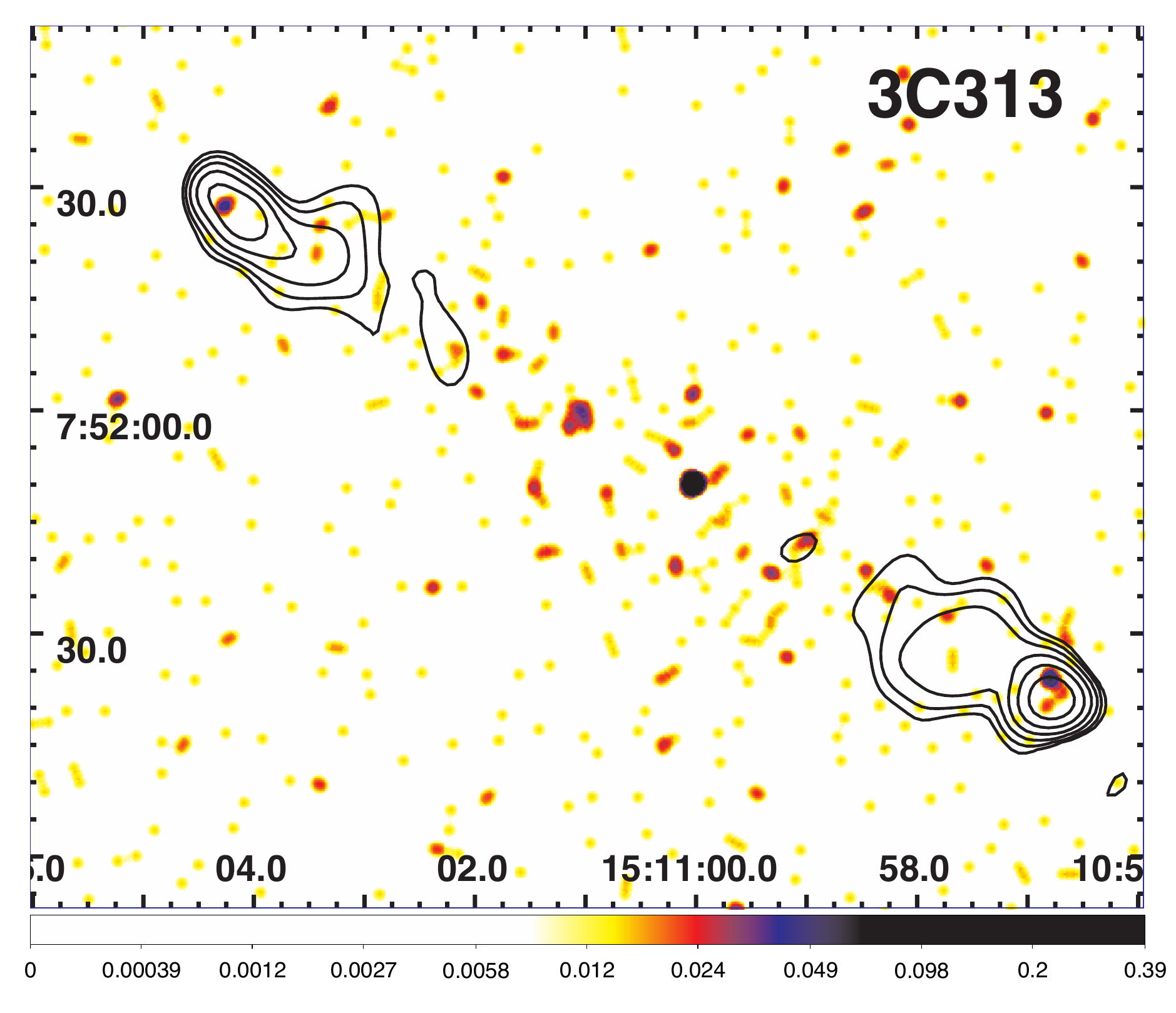}
\caption{The X-ray image of 3C\,313 for the energy band 0.5-7 keV.  The event
file has been regridded to a pixel size of 0.246\arcsec\ and
smoothed with a Gaussian of FWHM=2.0\arcsec.  The radio
contours (black) come from a 1.4 GHz map constructed from archival VLA
data and start at 15 mJy/beam, increasing by factors of two.  The
clean beam is 5\arcsec.
The VLA image at 1.4 GHz have been obtained via AIPS standard reduction procedure (http://www.aips.nrao.edu/cook.html).}
\label{fig:3c313app}
\end{figure}

\begin{figure}
\includegraphics[keepaspectratio=true,scale=0.90]{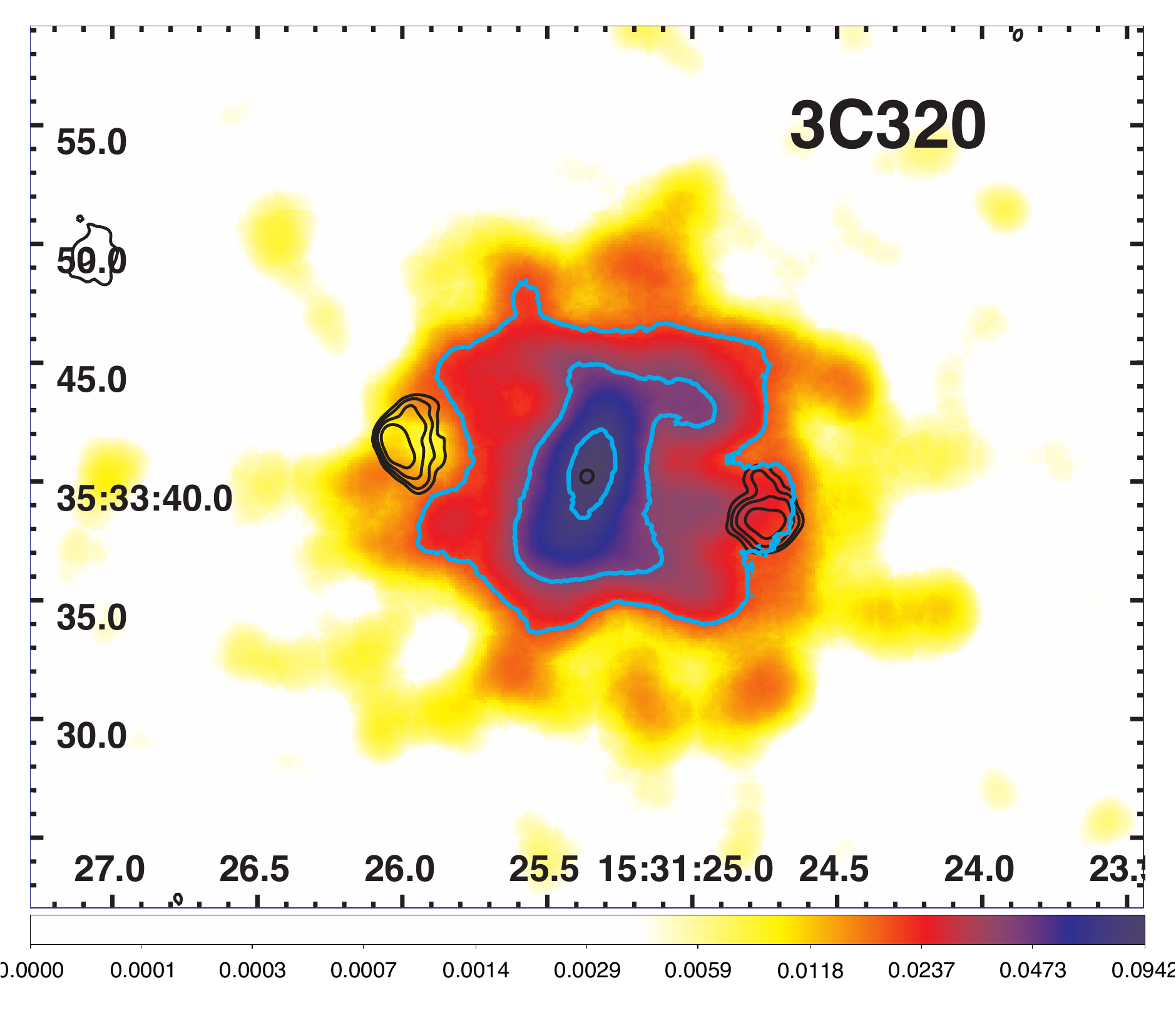}
\caption{The X-ray image of 3C\,320 for the energy band 0.5-7 keV.  The event
file has been regridded to a pixel size of 0.123\arcsec\ and
smoothed with a Gaussian of FWHM=2.5\arcsec.  X-ray contours
(white or cyan) start at 0.02 counts/pix and increase by factors of
two.  The radio contours (black) come from a 4.9 GHz map downloaded
from the NVAS and start at 0.5 mJy/beam, increasing by factors of two.
The clean beam is 0.40\arcsec\ x 0.36\arcsec\ with
major axis in PA=-47$^{\circ}$.}
\label{fig:3c320app}
\end{figure}

\begin{figure}
\includegraphics[keepaspectratio=true,scale=0.90]{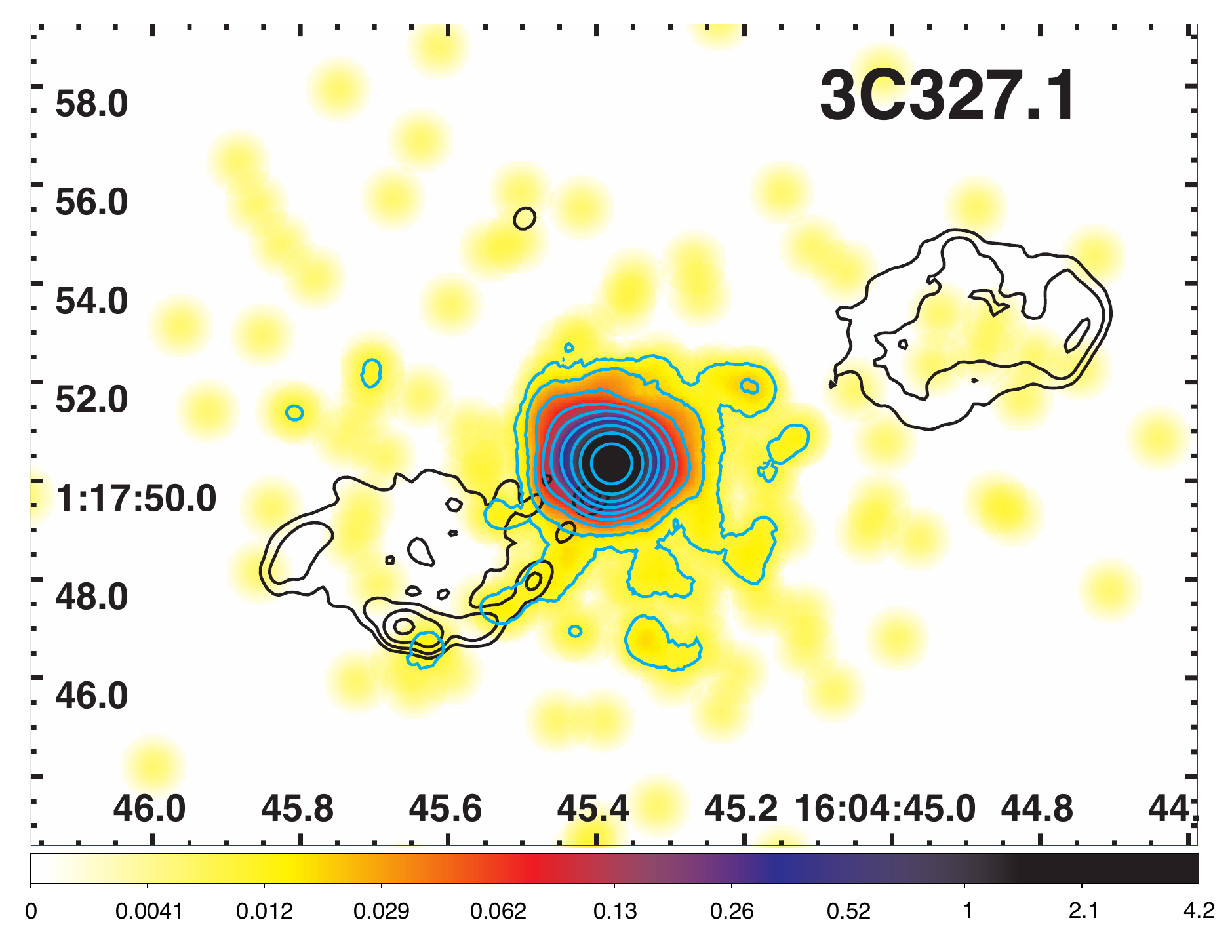}
\caption{The X-ray image of 3C\,327.1 for the energy band 0.5-7 keV.  The event
file has been regridded to a pixel size of 0.0615\arcsec\ and
smoothed with a Gaussian of FWHM=0.8\arcsec.  X-ray contours
(white or cyan) start at 0.01 counts/pix and increase by factors of
two.  The radio contours (black) come from a 4.9 GHz map kindly
supplied by R. Morganti and start at 1 mJy/beam, increasing by factors
of four.  The clean beam is
0.39\arcsec\ x0.36\arcsec\ with major axis in
PA=-38$^{\circ}$.  As indicated by the lowest X-ray contour, the S jet
seems to be marginally detected.}
\label{fig:3c327.1app}
\end{figure}

\begin{figure}
\includegraphics[keepaspectratio=true,scale=0.90]{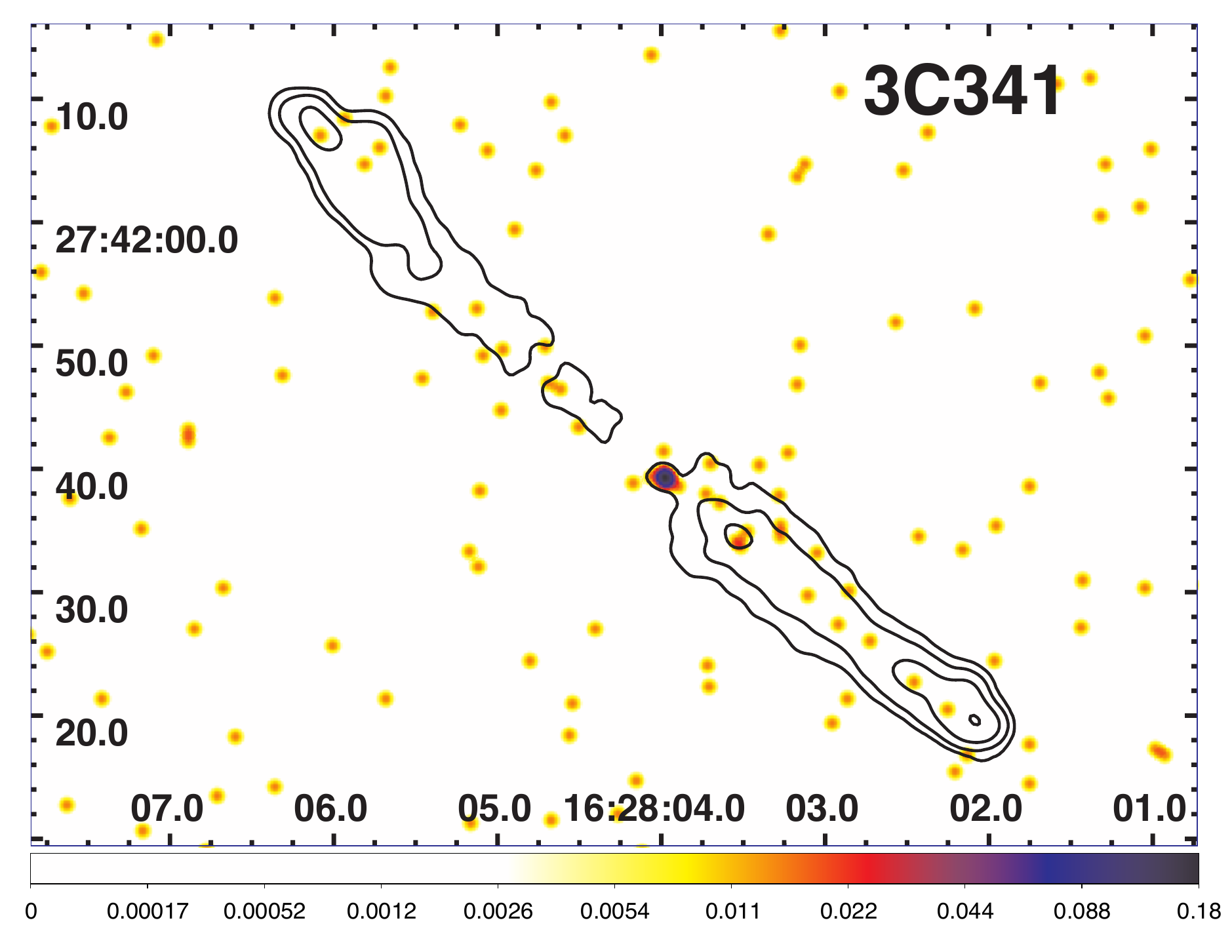}
\caption{The X-ray image of 3C\,341 for the energy band 0.5-7 keV.
  The event file has been regridded to a pixel size of
  0.123\arcsec\ and smoothed with a Gaussian of FWHM=1.0\arcsec.  The
  radio contours (black) come from an 8.5 GHz map kindly supplied by
  M. J. Hardcastle \citep{gilbert04} and start at 0.25 mJy/beam,
  increasing by factors of four.  The clean beam is 1.5\arcsec.}
\label{fig:3c341app}
\end{figure}

\begin{figure}
\includegraphics[keepaspectratio=true,scale=0.90]{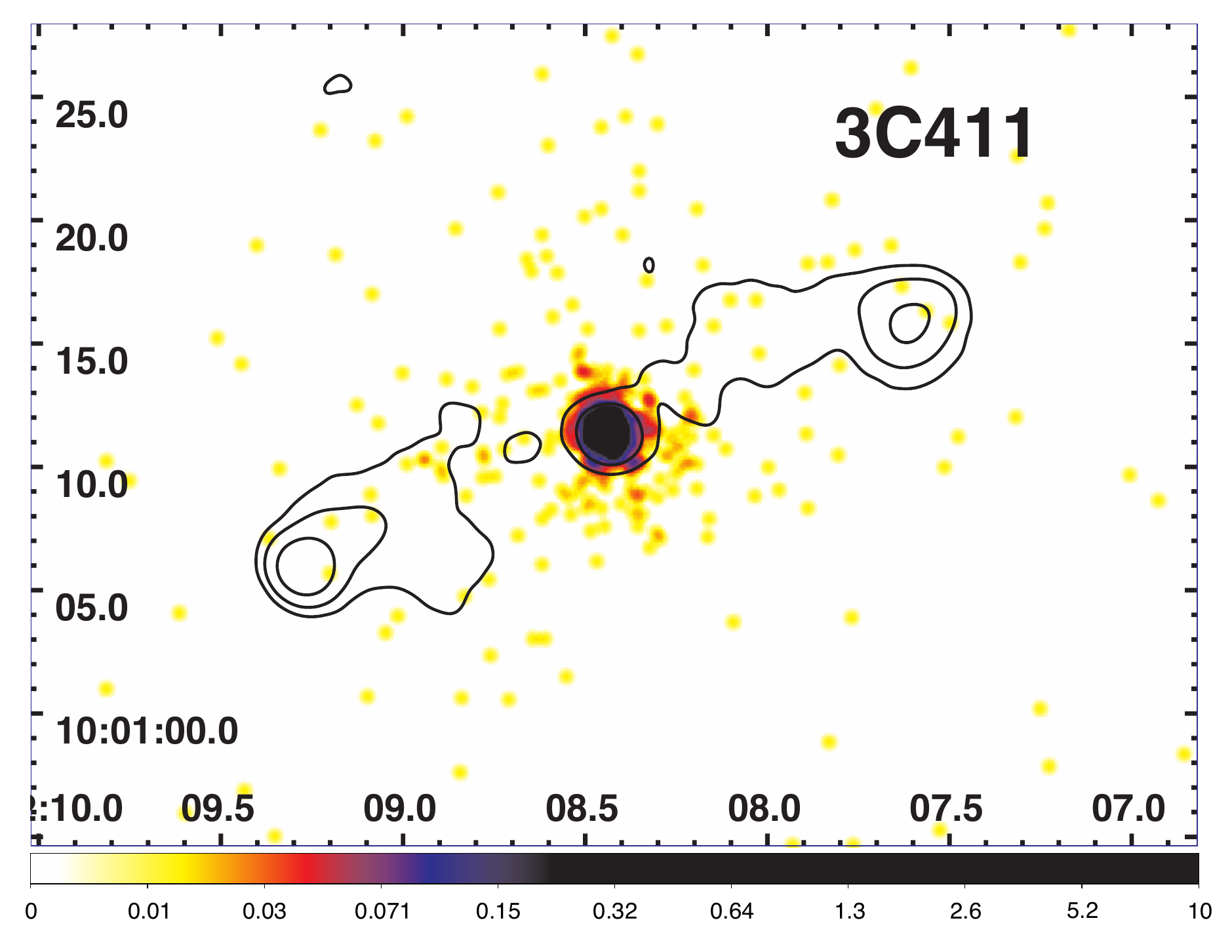}
\caption{The X-ray image of 3C\,411 for the energy band 0.5-7 keV.  The event
file has been regridded to a pixel size of 0.0615\arcsec\ and
smoothed with a Gaussian of FWHM=0.5\arcsec.  The radio
contours (black) come from a 4.9 GHz map downloaded from NED and
start at 4 mJy/beam, increasing by factors of four.  The clean beam is
1.6\arcsec\ x 1.5\arcsec\ with major axis in
PA=57$^{\circ}$.}
\label{fig:3c411app}
\end{figure}

\begin{figure}
\includegraphics[keepaspectratio=true,scale=0.90]{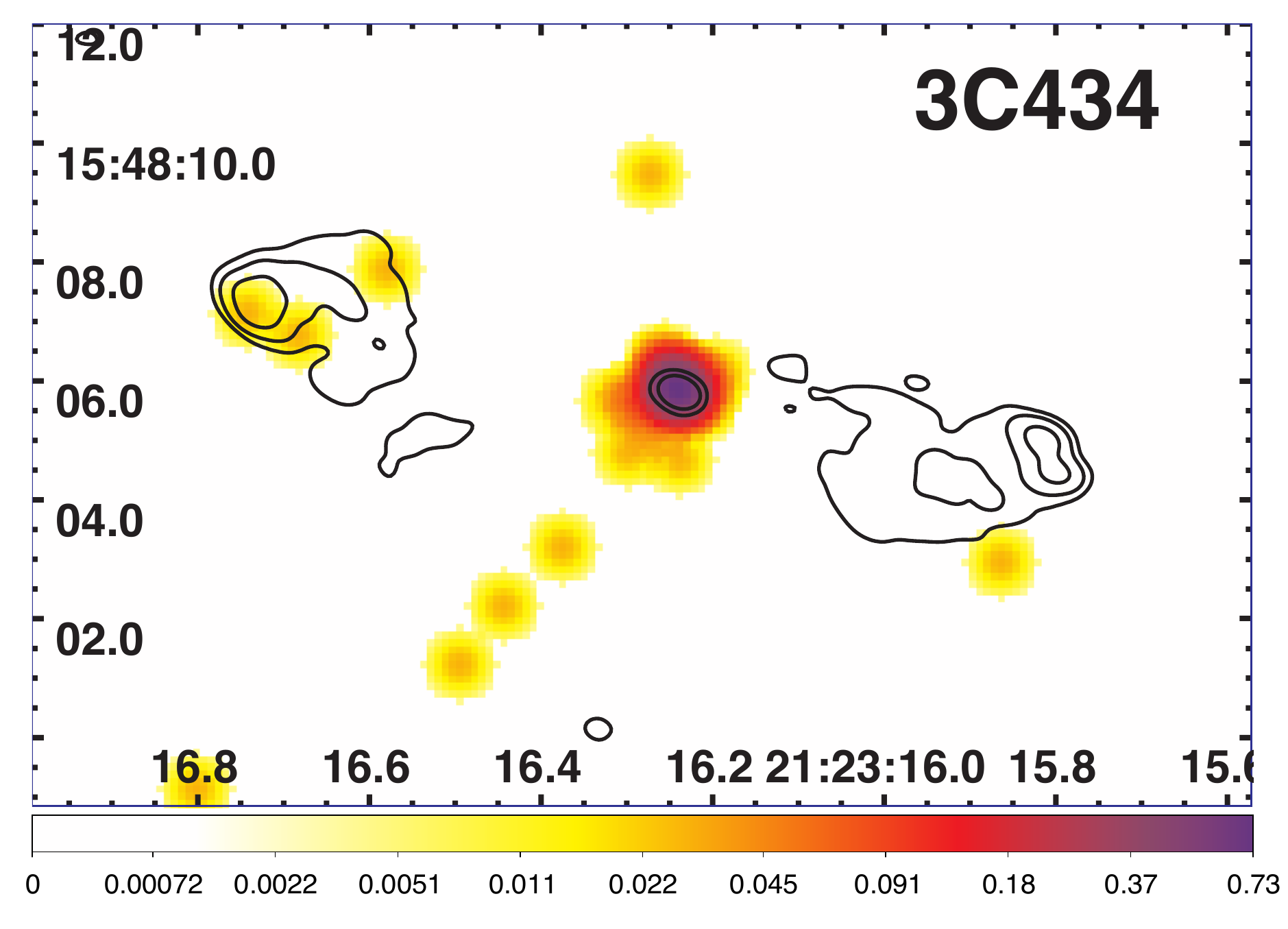}
\caption{The X-ray image of 3C\,434 for the energy band 0.5-7 keV.  The event
file has been regridded to a pixel size of 0.123\arcsec\ and
smoothed with a Gaussian of FWHM=0.72\arcsec.  The radio
contours (black) come from a 4.9 GHz map constructed from archival VLA
data and start at 0.5 mJy/beam, increasing by factors of four.  The
clean beam is 0.54\arcsec\ x 0.38\arcsec\ with major
axis in PA=66 $^{\circ}$.
The VLA image at 4.9 GHz have been obtained via AIPS standard reduction procedure (http://www.aips.nrao.edu/cook.html).}
\label{fig:3c434app}
\end{figure}

\begin{figure}
\includegraphics[keepaspectratio=true,scale=0.90]{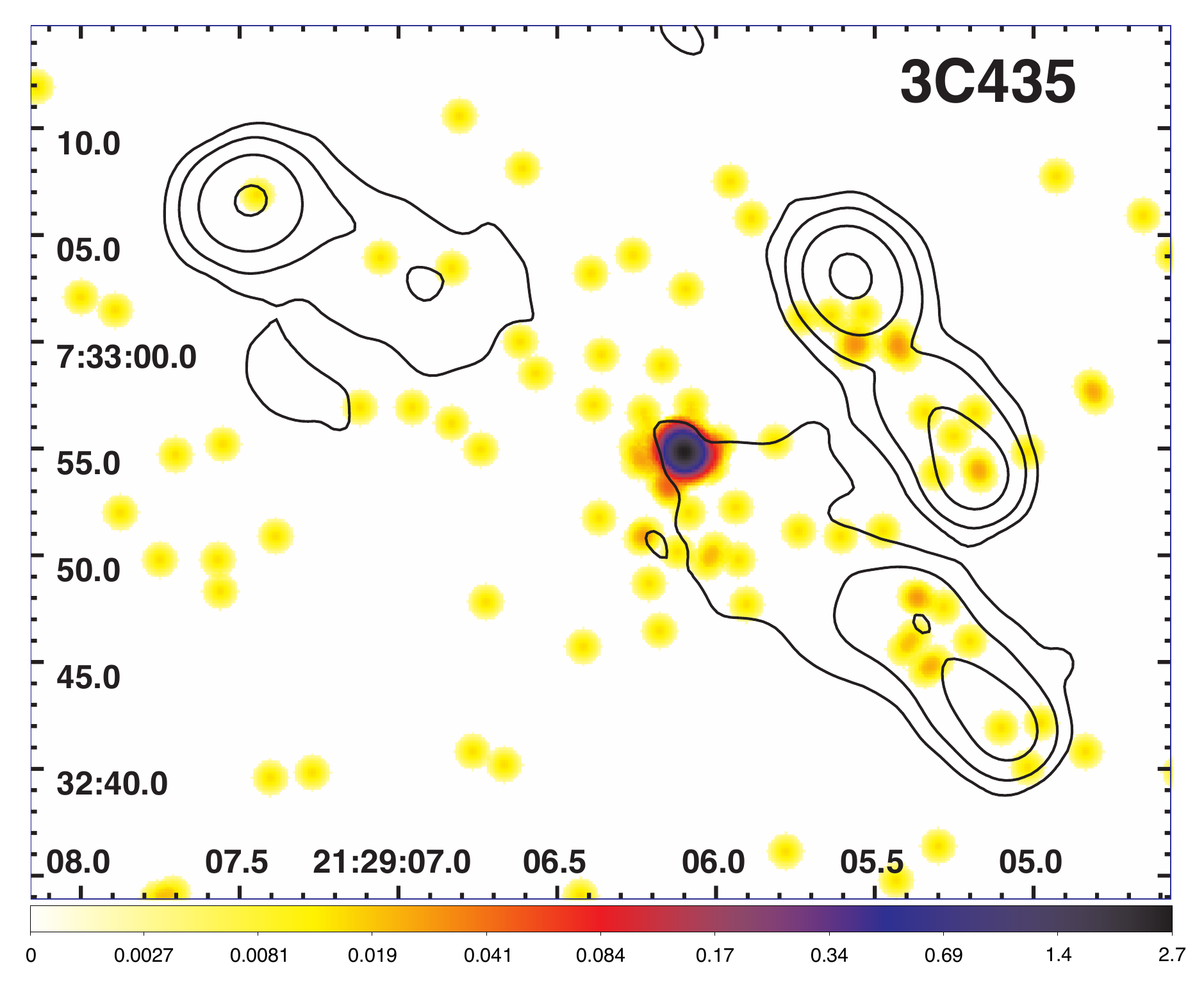}
\caption{The X-ray image of 3C\,435 for the energy band 0.5-7 keV.  The event
file has been regridded to a pixel size of 0.123\arcsec\ and
smoothed with a Gaussian of FWHM=1.0\arcsec.  The radio
contours (black) come from a 1.4 GHz map downloaded from NED and start
at 4 mJy/beam, increasing by factors of four.  The clean beam is
2.6\arcsec.  There are two FRII radio galaxies in
proximity \citep[see][for more detials]{mccarthy89}.  
X-ray emission was detected from the nucleus of the SE source (i.e., 3C\,435B, $\approx$300
counts), aligned with an unresolved radio component at 5 GHz.
The NW radio galaxy \citep[i.e., 3C\,435A][]{spinrad85},
is not detected in the \chn\ snapshot observation.}
\label{fig:3c435app}
\end{figure}

\end{document}